\documentclass[longauth]{aa}  

\usepackage{graphicx}
\usepackage{txfonts}
\usepackage{lipsum}
\usepackage{subcaption}       
                               
\usepackage{lscape}             
                               
\usepackage{placeins}           

\usepackage{hyperref}

\begin{document}
   \title{Horizontal magnetic fields dominate the weak-field quiet Sun in Sunrise III observations}

 \subtitle{Evidence from traditional and transformer-based inversions}

\author{
  Ryan~J.~Campbell\inst{1}\corrauth{ryan.campbell@qub.ac.uk} \and
  Carlos~Quintero~Noda\inst{2,3} \and
  Manuel~Collados\inst{2,3} \and
  Mihalis~Mathioudakis\inst{1} \and
  Andreas~Lagg\inst{4} \and
  Achim~Gandorfer\inst{4} \and
  Jose~Carlos~del~Toro~Iniesta\inst{5,6} \and
  Yukio~Katsukawa\inst{7,8,9} \and
  Pietro~Bernasconi\inst{10} \and
  Thomas~Berkefeld\inst{11} \and
  Alex~Feller\inst{4} \and
  Tino~L.~Riethmüller\inst{4} \and
  Alberto~Álvarez-Herrero\inst{12,6} \and
  Masahito~Kubo\inst{7,13} \and
  H.~N.~Smitha\inst{4} \and
  David~Orozco~Suárez\inst{5,6} \and
  Bianca~Grauf\inst{4} \and
  Michael~Carpenter\inst{10} \and
  Alexander~Bell\inst{11} \and
  Valentín~Martínez~Pillet\inst{2,6} \and
  Francisco~Javier~Bailén\inst{5,6} \and
  Julian~Blanco~Rodríguez\inst{14,6} \and
  Juan~Sebastián~Castellanos~Durán\inst{4} \and
  Edvarda~Harnes\inst{4} \and
  Johannes~Hölken\inst{4} \and
  Francisco~A.~Iglesias\inst{4,15} \and
  Ryohtaroh~T.~Ishikawa\inst{16} \and
  Yusuke~Kawabata\inst{7} \and
  Takuma~Matsumoto\inst{17} \and
  Takayoshi~Oba\inst{18,4} \and
  Azaymi~L.~Siu-Tapia\inst{5,6} \and
  Hanna~Strecker\inst{5,6} \and
  Dušan~Vukadinović\inst{19,4} \and
  Hirohisa~Hara\inst{7} \and
  Toshifumi~Shimizu\inst{20,21}
}

\institute{
  Astrophysics Research Centre, Queen's University Belfast, Belfast, Northern Ireland, BT7 1NN, United Kingdom \and
  Instituto de Astrofísica de Canarias, Vía Láctea, s/n, E-38205 La Laguna, Spain \and
  Departamento de Astrof\'isica, Universidad de La Laguna, E-38206 La Laguna, Tenerife, Spain \and
  Max-Planck-Institut für Sonnensystemforschung, Justus-von-Liebig-Weg 3, 37077 Göttingen, Germany \and
  Instituto de Astrofísica de Andalucía, CSIC, Glorieta de la Astronomía s/n, 18008 Granada, Spain \and
  Spanish Space Solar Physics Consortium \and
  National Astronomical Observatory of Japan, 2-21-1 Osawa, Mitaka, Tokyo 181-8588, Japan \and
  Department of Astronomy, The University of Tokyo, 7-3-1, Hongo, Bunkyo-ku, Tokyo 113-0033, Japan \and
  Department of Astronomical Science, The Graduate University for Advanced Studies (SOKENDAI), 2-21-1 Osawa, Mitaka, Tokyo 1818588, Japan \and
  Johns Hopkins University Applied Physics Laboratory, 11100 Johns Hopkins Road, Laurel, Maryland, USA \and
  Institut für Sonnenphysik (KIS), Georges-Köhler-Allee 401a, 79110 Freiburg, Germany \and
  Instituto Nacional de T\'ecnica Aeroespacial (INTA), Ctra. de Ajalvir, km. 4, E-28850 Torrejón de Ardoz, Spain \and
  Department of Astronomical Science, The Graduate University for Advanced Studies (SOKENDAI), 2-21-1 Osawa, Mitaka, Tokyo 181-8588, Japan \and
  Universitat de Valencia Catedrático José Beltrán 2, E-46980 Paterna-Valencia, Spain \and
  Grupo de Estudios en Heliofísica de Mendoza, CONICET, Universidad de Mendoza, Boulogne sur Mer 683, 5500 Mendoza, Argentina \and
  National Institute for Fusion Science, 322-6 Oroshi-cho, Toki City 509-5292, Japan \and
  Centre for Integrated Data Science, Institute for Space-Earth Environmental Research, Nagoya University, Furocho, Chikusa-ku, Nagoya, Aichi 464-8601, Japan \and
  Advanced Research Center for Space Science and Technology, Institute of Science and Engineering, Kanazawa University, Kakuma-machi, Kanazawa, Ishikawa 920-1192, Japan \and
  Institut für Physik, Universität Graz, Universitätsplatz 5, 8010 Graz, Austria \and
  Department of Earth and Planetary Science, The University of Tokyo, 7-3-1, Hongo, Bunkyo-ku, Tokyo 113-0033, Japan \and
  Institute of Space and Astronautical Science, Japan Aerospace Exploration Agency, 3-1-1, Yoshinodai, Chuo-ku, Sagamihara, Kanagawa 252-5210, Japan
}

\date{Received August 03, 2026, Accepted October 26, 2026}

\abstract
{The relative prevalence of horizontal and vertical magnetic fields in the quiet-Sun remains debated, owing to the weak linear polarisation signals that make the inferred magnetic inclination distributions sensitive to observational quality and inversion methodology.}
{We investigate the magnetic topology of the quiet-Sun in {\sc Sunrise~iii} observations by independently applying transformer-based and traditional inversion techniques to the same spectropolarimetric dataset.}
{We analyse observations obtained with the Sunrise Chromospheric Infrared spectroPolarimeter (SCIP). Transformer-based inversions (SINN), trained exclusively on synthetic Stokes profiles generated from a large radiative magnetohydrodynamic simulation, are compared with independent inversions performed using the \textsc{DeSIRe} code. We separate the selected pixels into weak- and strong-field populations according to their circular-polarisation amplitudes.}
{Linear polarisation exceeding $4\sigma$ is detected in $34.6\%$ of the field of view, enabling the magnetic topology to be constrained over a substantially larger detected population than in typical short-integration Hinode observations. Both inversion methods recover a quiet-Sun magnetic topology in which the weak fields are strongly dominated by the horizontal component, whereas strong fields exhibit a substantially more balanced topology. At $\log\tau=-1$, the median horizontal and unsigned line-of-sight field components in the weak-field population are 63 and 7~G with \textsc{DeSIRe}, and 60 and 6.5~G with SINN, respectively, corresponding to a horizontal-to-line-of-sight ratio of approximately nine for both methods.}
{The agreement between the two fundamentally different optimisation methodologies indicates that the weak-field quiet-Sun in this dataset is characterised by a substantially stronger horizontal than vertical magnetic field component. These results demonstrate that transformer-based inversions can be successfully transferred from synthetic training data to real spectropolarimetric observations, while providing inference orders of magnitude faster than traditional inversion techniques.}


\maketitle
\nolinenumbers

\section{Introduction}
The angular distribution of internetwork magnetic fields constrains how quiet-Sun magnetism is generated and reorganised by convection. Predominantly vertical fields would emphasise concentration and transport in intergranular lanes, whereas an enhanced horizontal component may reflect small-loop emergence, convective reorientation, or magnetic expansion with height. Determining this distribution therefore helps to test models of small-scale dynamo action and flux emergence and constrains magnetic-energy transport through the photosphere \citep{2019LRSP...16....1B}.  Hinode \citep{kosugi2007SoPh..243....3K, lites2013SoPh..283..579L} observations suggested that internetwork magnetic fields may be predominantly horizontal \citep{Lites2008, Suarez2012, Lites2017}. Analyses employing spectral lines formed deeper in the photosphere have reported a more balanced or vertical topology \citep{campbell2021A&A...647A.182C,Campbell2023}. Despite nearly two decades of Hinode observations, the angular distribution of internetwork magnetic fields remains debated \citep{andres2009,borrero2011,borrero2012,danilovic2016}. Observational noise and inversion degeneracies can substantially influence inferred inclination distributions, motivating new measurements with improved spatial resolution and polarimetric sensitivity \citep{andres2014}. Indeed, the properties of so-called linear polarisation features were found to depend on the treatment of noise in previous \textsc{Sunrise} observations \citep{kianfar2018SoPh..293..123K}.

Resolving this question requires both observations capable of detecting weak linear polarisation signals and inversion techniques that recover reliable atmospheric stratifications from those measurements. Inversion codes have become indispensable tools for interpreting solar spectropolarimetric data \citep{inversions}. Despite their success, traditional inversion methods remain computationally expensive. Modern solar facilities routinely produce datasets containing millions of spectra, while forthcoming observations from facilities such as the Daniel K. Inouye Solar Telescope and the European Solar Telescope will further increase both the volume and complexity of available data \citep{DKIST, EST}. This motivates the development of alternative or complementary approaches capable of providing atmospheric inference at substantially higher throughput.

Machine-learning (ML) techniques have emerged as a promising route toward rapid spectropolarimetric inversion. Early studies demonstrated that artificial neural networks could learn direct mappings between Stokes spectra and atmospheric parameters, enabling inference orders of magnitude faster than conventional inversion techniques \citep{RamosLRSP}. Increasingly sophisticated ML architectures, including convolutional neural networks and other deep-learning approaches, have been applied successfully to solar spectropolarimetric problems \citep{Ramos2019, Ramos2025, Navarro}. These methods offer the possibility of near-instantaneous atmospheric inference once training has been completed. A fundamental limitation of supervised ML inversions, however, is that they require labelled training data. Concerns have  been raised regarding the extent to which models trained on simulations can be expected to generalise to real observations, especially in cases where the model may need to extrapolate beyond its training data \citep{yang2025ApJ...995..146Y}. Demonstrating successful transfer from synthetic training data to real observations therefore remains one of the most important outstanding challenges for ML-based inversion techniques.

Transformer architectures have recently transformed many areas of ML through their use of self-attention mechanisms that efficiently capture long-range relationships within sequential data \citep{vaswani2017attention}. \citet{Campbell2026} introduced the Stokes Inversion based on Neural Networks (SINN) framework, a transformer-based inversion architecture designed to infer atmospheric stratifications directly from spectropolarimetric observations. Using synthetic datasets, SINN was shown to outperform multilayer perceptron (MLP) architectures and to recover more physically regularised atmospheric stratifications. Transformers have also been used to frame the inversion as a translation task \cite{ramos_jaime_2025}.

The observations obtained by the Sunrise Chromospheric Infrared Spectro-Polarimeter (SCIP; \citealp{2026arXiv260317929K}) aboard the {\sc Sunrise~iii} balloon-borne solar observatory \citep{sunrise3} provide an opportunity to revisit  quiet-Sun magnetic topology using substantially improved spectropolarimetric observations while simultaneously testing whether transformer-based inversions trained exclusively on synthetic data transfer successfully to real observations. Compared with previous space-based observations, SCIP has revealed abundant linear-polarisation signals without the need for the very long temporal integrations often employed in earlier studies. In this work we focus on observations of the magnetically sensitive Fe~\textsc{i}~8468~\AA\ line obtained in a quiet-Sun region. Independent atmospheric inversions of the SCIP observations have already been performed using \textsc{DeSIRe} \citep{desire}, providing a physically motivated reference against which the transformer predictions may be compared \citep{Campbell2026Network}. Consequently, these observations not only permit an assessment of whether SINN can be transferred successfully to real spectropolarimetric data, but also provide a stringent test of whether transformer-based inversions recover the same physical conclusions regarding quiet-Sun magnetic topology as a mature inversion framework.

In this work, we train SINN exclusively on synthetic spectropolarimetric observations generated from radiative magnetohydrodynamic simulations and evaluate its performance on both synthetic and observational datasets. We first assess the ability of the model to generalise to an independent simulation snapshot not used during training. We then deploy the trained model on quiet-Sun SCIP observations and compare the inferred atmospheric structure with independent \textsc{DeSIRe} inversions. We focus on whether the independently inferred magnetic topology is consistent between the two different inversion methodologies.

\section{Data}

\subsection{Simulations}

We use snapshots from a radiative MHD simulation produced with MURaM \citep{rempel}. The snapshot was selected from the SPIN4D MURaM archive \citep{SPIN4D}, specifically from the \texttt{SPIN4D\_SSD\_Large} case. This small-scale dynamo simulation spans a very large horizontal domain ($50\times50$~Mm) while retaining the high spatial resolution ($16$~km in the horizontal direction), allowing extended quiet Sun network structures to develop self-consistently within realistic magnetoconvection. The simulation we use had a purely vertical magnetic field of $+200$~G, $-150$~G, and $-50$~G added in three quadrants. Full details of the simulations are available in \cite{SPIN4D}. The two snapshots we use for training (and validation) and generalisation testing are 132 minutes apart. This interval is long compared with the photospheric granular evolution time, so the generalisation snapshot is not expected to retain pixel-to-pixel spatial correlations with the training snapshot.

\subsection{Observations}
The data analysed here were obtained with the SCIP slit spectropolarimeter aboard Sunrise III on 11 July 2024 between 04:28:47--06:15:55 UT. The scanned quiet Sun region was at disk centre. The instrument samples the Fe I 8468\,\AA\ line, providing full Stokes measurements over a two-dimensional field of view constructed via slit scanning. Standard data reduction \citep{solanki2026} and preprocessing steps were applied, including continuum normalisation. Spectral windows were defined to isolate the Fe I 8468\,\AA\ line core and continuum regions in the channel SP1. The standard deviation, $\sigma$, in continuum regions of Stokes~$Q$ and $V$ was $4.68\times10^{-4}\,I_\mathrm{c}$ and $4.16\times10^{-4}\,I_\mathrm{c}$, respectively, where $I_\mathrm{c}$ is the spatially averaged continuum intensity. More details of the observations are available in \cite{Campbell2026Network}. 

The SCIP dataset is particularly well suited to the analysis presented in this work because it combines high spatial resolution and high polarimetric sensitivity observations in the magnetically sensitive Fe~\textsc{i}~8468~\AA\ line. Consequently, the observations simultaneously capture both strong quiet-Sun (network) concentrations and weak (internetwork) magnetic fields, providing a stringent test of the transformer inversion across a broad range of magnetic environments. At the same time, the coexistence of strong (network) and weak (internetwork) structure broadly resembles the magnetic morphology present in the MURaM simulations used to generate the synthetic training data, making the observations well suited for evaluating synthetic-to-observation transfer.

\section{Methods}

\subsection{Training strategy}

The transformer was trained entirely on synthetic spectropolarimetric observations generated from MURaM simulations. Each wavelength sample was treated as an individual token, represented by the corresponding four-component Stokes vector $(I,Q,U,V)$. The encoder employed self-attention across the spectral dimension, allowing information from all wavelength positions to interact directly and enabling the neural network to identify correlations between widely separated spectral features. The decoder consisted of a set of learned optical depth queries representing the atmospheric layers to be inferred. Cross-attention between the decoder queries and the encoded spectral representation allowed each depth point to selectively attend to the portions of the spectrum most relevant to its prediction, while self-attention within the decoder enabled information exchange between neighbouring atmospheric layers and promoted vertically coherent stratifications. A more detailed description of this Stokes Inversion based on Neural Networks (SINN) architecture can be found in \citet{Campbell2026} and the code is publicly available\footnote{https://github.com/r-j-campbell/SINN-inversions/}.

\begin{figure*}
    \centering
    \includegraphics[width=\linewidth]{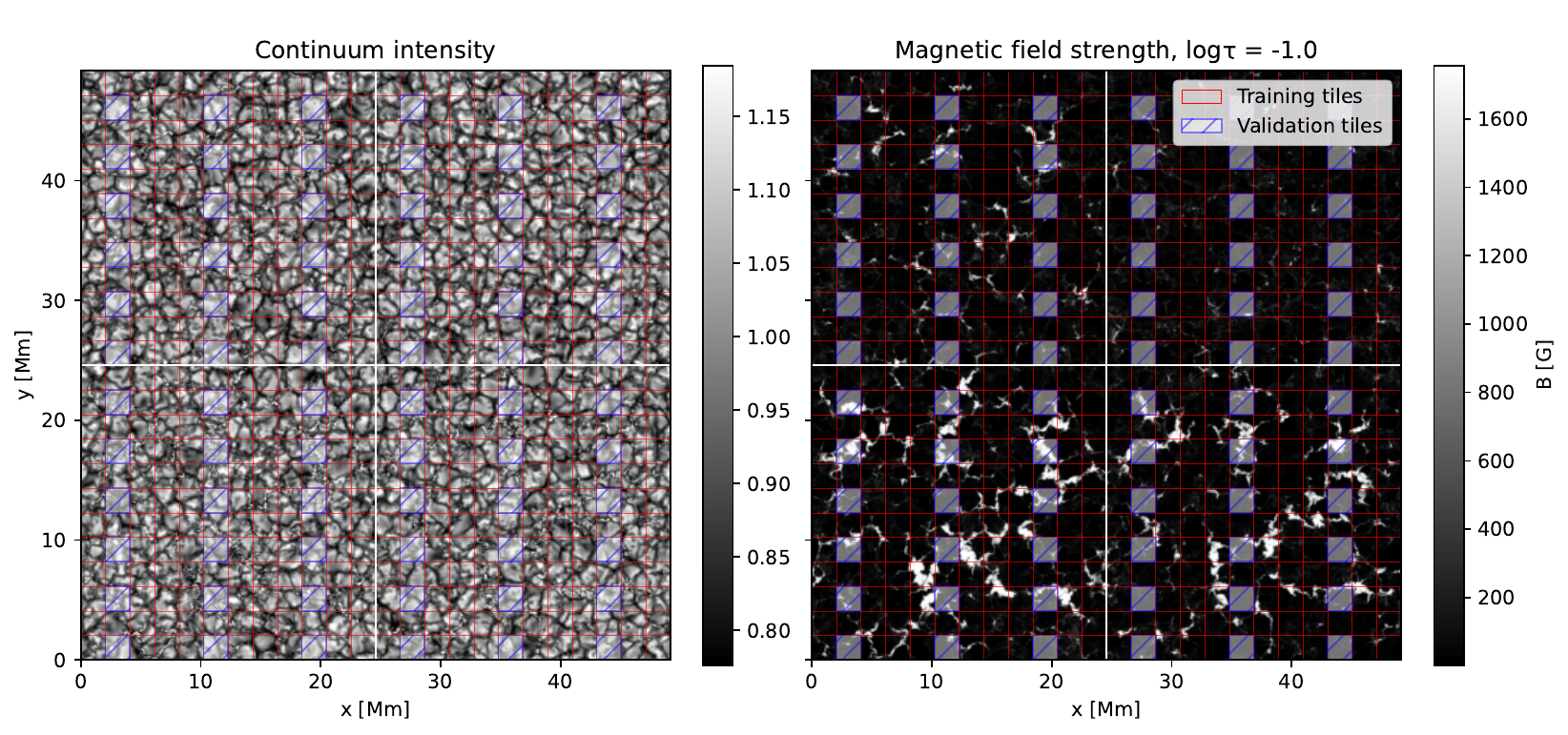}
    \caption{Continuum intensity (left) and magnetic field strength at $\log\tau=-1$ (right) from the MURaM simulation used to generate the synthetic training data. The simulation domain was divided into non-overlapping $128\times128$ pixel tiles ($2.05\times2.05$ Mm). Tiles assigned to the validation set are indicated by the shaded overlays, while the remaining tiles were used for training. This deterministic tile-based partitioning ensures complete spatial separation between training and validation samples, preventing information leakage between neighbouring pixels and providing a more stringent optimisation procedure than a random pixel-wise split. The resulting split produced approximately $8.3\times10^6$ training pixels and $1.2\times10^6$ validation pixels prior to epoch-wise subsampling.}
    \label{fig:training_pattern}
\end{figure*}

\begin{figure*}
\centering
    \includegraphics[width=.9\textwidth]{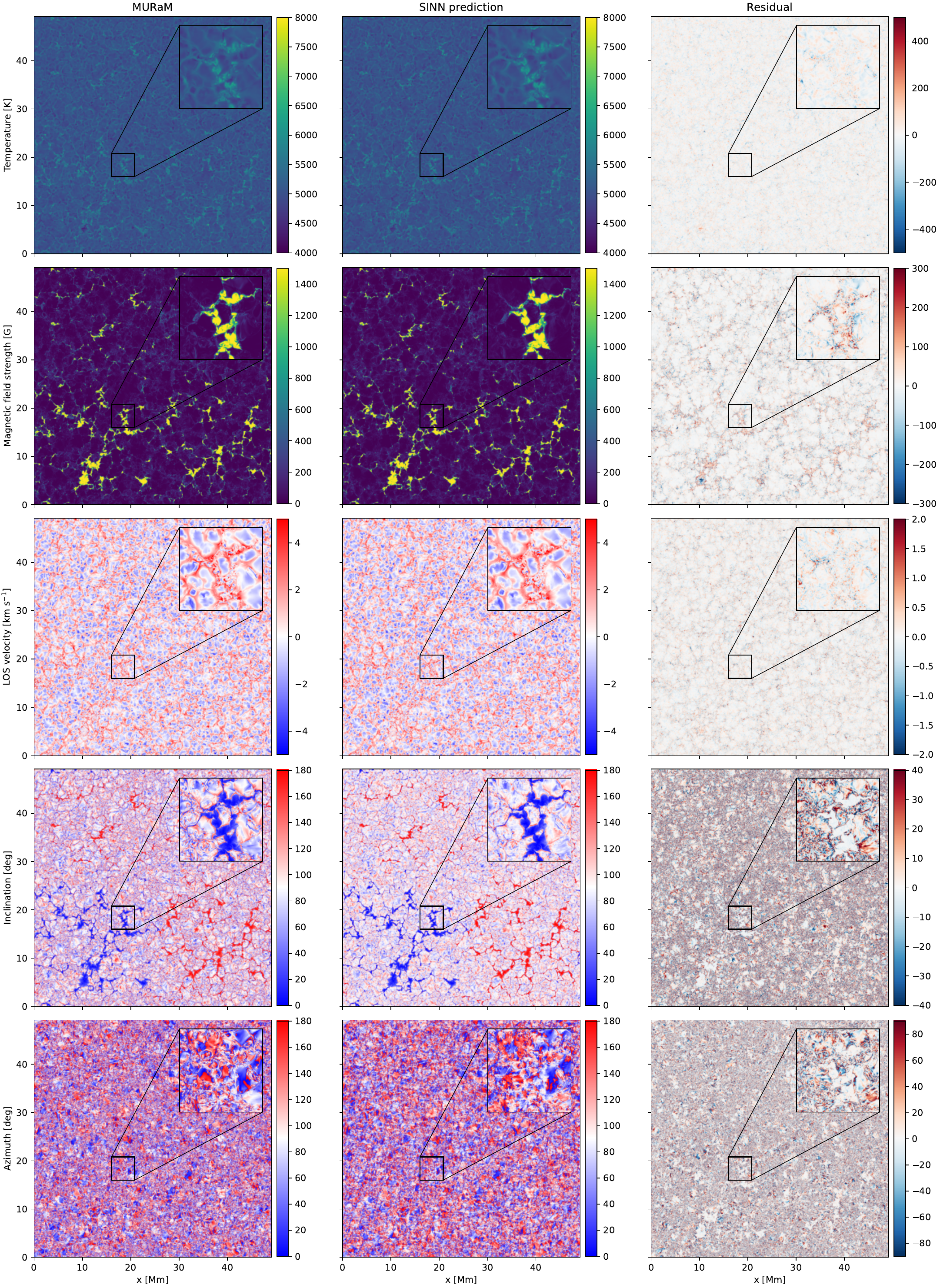}
    \caption{Maps of atmospheric parameters predicted by SINN on an independent MURaM snapshot not used during training. Atmospheric maps at $\log\tau=-1$ are shown for temperature, magnetic field strength, line-of-sight velocity, magnetic inclination, and magnetic azimuth. The left column shows the simulation atmosphere (ground truth), the middle column shows the SINN predictions, and the right column shows the residuals. Insets indicate a region of interest used to compare fine-scale structure.}\label{fig:generalisation}
\end{figure*}

The atmospheric quantities of interest were temperature, magnetic field strength, $B$, line-of-sight velocity, magnetic inclination, $\gamma$, and magnetic azimuth, $\phi$, as functions of optical depth. To avoid discontinuities associated with the 180° ambiguity of the magnetic azimuth, the network was trained to predict $\sin(2\phi)$ and $\cos(2\phi)$ rather than $\phi$ directly. Atmospheric parameters were predicted on a uniform optical depth grid spanning $\log\tau = -3.0$ to $0.5$ in steps of $\log\tau = 0.1$.

The synthetic training data were derived from a large MURaM snapshot containing over nine million spatial pixels. \textsc{DeSIRe} was used to synthesise the spectral line profiles on a dense 5~m$\mathrm{\AA}$ sampling, enabling re-sampling to match the SCIP observations when required prior to training. To reduce memory requirements while preserving the full diversity of magnetic structures present in the simulation, a subset of $4.5\times10^{6}$ training profiles was randomly sampled during each training epoch. A key concern when training on spatially structured simulations is information leakage between the training and validation sets. Neighbouring pixels within MHD simulations are strongly correlated, and random pixel-wise splitting could therefore lead to overly optimistic estimates of predictive performance. To mitigate this effect, the simulation domain was partitioned into non-overlapping $128\times128$ pixel tiles. Entire tiles were assigned either to the training set or the validation set using the deterministic spatial pattern shown in Fig.~\ref{fig:training_pattern}. Consequently, all pixels within a validation tile were excluded from training, ensuring that the network was evaluated on spatial regions that were not encountered during hyperparameter optimisation.

Input Stokes spectra were standardised using global per-Stokes scaling factors computed from the synthetic dataset. The mean and standard deviation were calculated independently for Stokes $I$, $Q$, $U$, and $V$ across the training cube and subsequently applied to all training, validation, and inference data. Atmospheric targets were standardised independently at each optical depth point to account for the large dynamic range present in thermodynamic and magnetic quantities throughout the atmosphere. 

Model optimisation employed the AdamW algorithm with a mean-squared-error (MSE) loss function computed across all predicted atmospheric parameters and optical depth layers. Hyperparameter exploration was performed over learning rate, hidden dimension, and attention-head configurations. For each configuration, training proceeded until the validation loss failed to improve for five consecutive epochs, at which point early stopping was triggered and the parameters corresponding to the lowest validation loss were retained.

Before deploying on real observational data, the synthetic spectra were resampled onto the wavelength grid of the target instrument. To improve robustness against uncertainty in the observational noise properties, additive Gaussian noise was applied dynamically during training. For each presentation of a training profile, the noise standard deviation was drawn randomly from a uniform range encompassing the estimated polarimetric noise level of the observations, and a new noise realisation was generated. Consequently, the same atmospheric model was observed under different noise conditions throughout training. This strategy discourages the network from adapting to a specific noise pattern or precise noise level and instead promotes the learning of spectral features that remain stable across a realistic range of observing conditions.

\subsection{DeSIRe inversions}
As an independent benchmark, we compare the transformer predictions with atmospheric parameters inferred from the same observations using the \textsc{DeSIRe} inversion code \citep{desire}. The \textsc{DeSIRe} inversions are the same as those used by \cite{Campbell2026Network}. The SCIP observations were inverted assuming a single magnetic atmospheric component per pixel. Photospheric Fe~\textsc{i} lines were treated in local thermodynamic equilibrium (LTE), while the chromospheric Ca~\textsc{ii} lines were treated in Non-LTE. Atmospheric parameters were allowed to vary with optical depth, including temperature, magnetic field strength, inclination, azimuth, and line-of-sight velocity. To reduce sensitivity to the initial atmospheric model, each pixel was inverted ten times using different initialisations and the minimum-$\chi^2$ solution was retained. Although the \textsc{DeSIRe} inversions utilised the full SCIP spectral dataset, including the Ca~\textsc{ii} lines, the transformer was trained exclusively on Fe~\textsc{i}~8468~\AA\ profiles and predicts atmospheric parameters only over the photospheric optical-depth range considered during training. Our comparison is therefore restricted to optical depths where the Fe~\textsc{i} line retains significant sensitivity and where both approaches are expected to provide reliable constraints. Because no atmospheric ground truth exists for the SCIP observations, the comparison between SINN and \textsc{DeSIRe} is intended to assess the consistency of the physical inferences obtained by two different inversion methodologies rather than to establish the superiority of either approach. Finally, we note that we also processed \textsc{DeSIRe} inversions of only the Fe~\textsc{i}~8468~\AA\ line and found that it did not meaningfully change the results presented in this manuscript.

\section{Results}
\subsection{Validation of SINN on synthetic data}
Before applying SINN to observations, we assessed its ability to generalise to an independent snapshot from the same MURaM simulation that was not used during training. The transformer was trained exclusively on synthetic spectra generated from one independent snapshot and evaluated on a temporally evolved snapshot separated by $132$~minutes. Although both snapshots originate from the same MURaM simulation, they contain distinct granular, intergranular, and magnetic structures, providing a straightforward test of whether the network has learned a physically meaningful mapping between Stokes spectra and atmospheric parameters. The evaluation snapshot was processed using the same synthesis and preprocessing pipeline as the training data. Predicted atmospheric stratifications were compared directly with the corresponding simulation atmosphere over the optical-depth range considered during training.

Figure \ref{fig:generalisation} compares representative atmospheric maps from the independent test snapshot with the corresponding transformer predictions at $\log\tau=-1$. The model successfully reproduces the large-scale thermodynamic and magnetic structure of the atmosphere, including granular temperature variations, magnetic flux concentrations, velocity patterns, magnetic polarity, and coherent azimuthal structure. Fine-scale differences remain present, but the overall agreement indicates that the network has learned physically meaningful relationships between the synthetic spectra and the underlying atmospheric state.

\begin{figure*}
    \includegraphics[width=\textwidth]{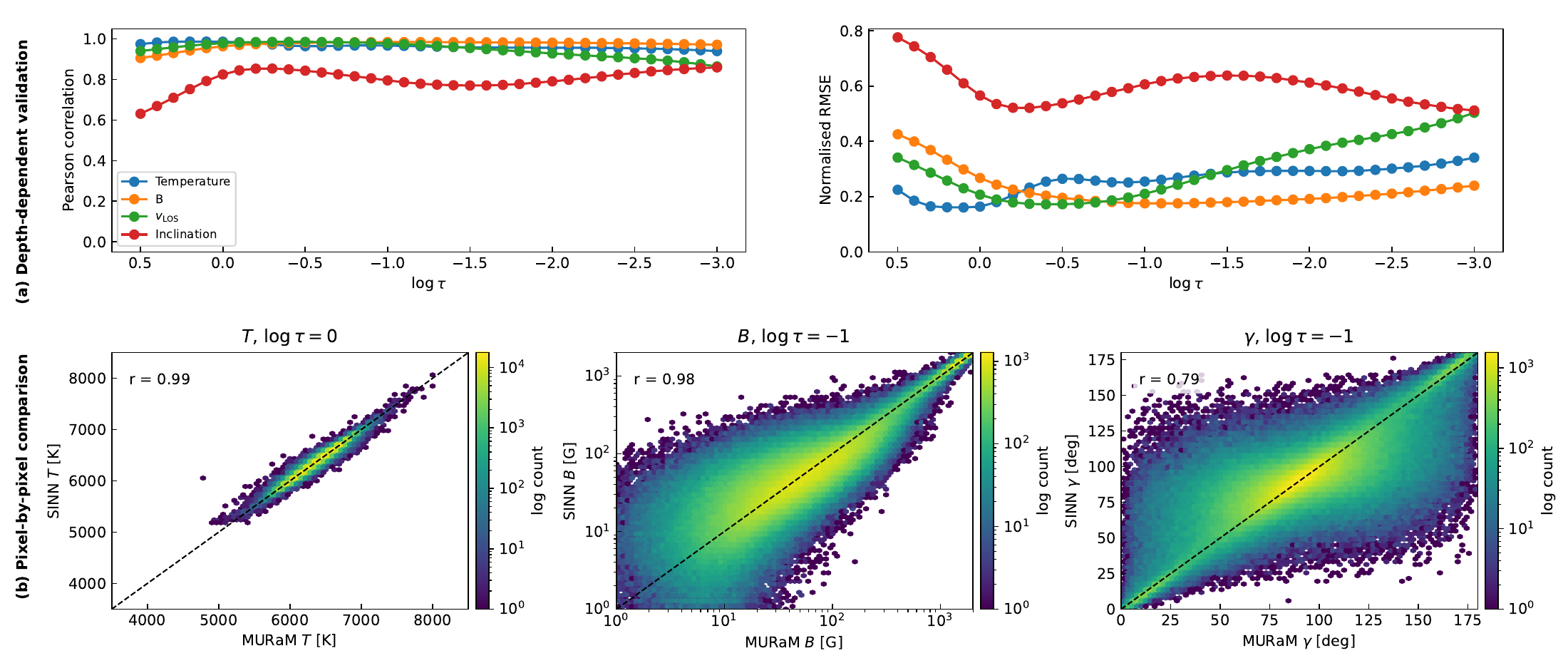}
    \caption{Generalisation testing of SINN on an independent MURaM snapshot not used during training. (a) Depth-dependent performance of the model, showing Pearson correlation coefficients (left) and normalised root-mean-square errors (right) as functions of optical depth for temperature, magnetic field strength, line-of-sight velocity, and magnetic inclination. The RMSE was normalised by the standard deviation of the corresponding ground-truth parameter distribution at each optical depth. (b) Pixel-by-pixel comparisons between the SINN predictions and the ground-truth atmospheric parameters for $1$~million randomly sampled pixels, for temperature at $\log\tau=0$, magnetic field strength at $\log\tau=-1$, and magnetic inclination at $\log\tau=-1$. The colour scale represents the logarithm of the number of pixels in each bin, while the dashed line indicates one-to-one agreement. Pearson correlation coefficients for the pixel-by-pixel comparisons are shown in the upper-left corner of each panel. }\label{fig:metrics}
\end{figure*}

To quantify predictive performance throughout the atmosphere, we computed validation metrics as a function of optical depth. Figure~\ref{fig:metrics} shows the resulting correlations and prediction errors for all inferred atmospheric quantities. The highest accuracy is generally achieved in the $\log\tau=-1$ to $\log\tau=0.0$ region, where the spectral lines exhibit their strongest sensitivity, while performance gradually decreases toward the uppermost and deepest layers. Nevertheless, the model maintains strong correlations across the full depth range considered in this work. The $\gamma$ has two regions of maximum correlation (and minimum RMSE): one in the deep photosphere and one in the upper photosphere.

Figure~\ref{fig:metrics} also compares the atmospheric parameters inferred by SINN against the ground truth MURaM simulation for temperature, magnetic field strength, and magnetic inclination on a pixel-by-pixel basis. The distributions are strongly correlated, with the highest agreement obtained for temperature and somewhat larger scatter present in the magnetic quantities. The similarity between the ground-truth and SINN distributions, together with the strong pixel-by-pixel correlations, indicate that the transformer generalises successfully to an unseen MURaM atmosphere and therefore provides a suitable basis for deployment on observational data.

\subsection{Deployment of SINN on Sunrise observations}

Having established that the model generalises to an independent simulation snapshot, we next investigate whether the learned mapping transfers to real spectropolarimetric observations obtained with {\sc Sunrise~iii}/SCIP. 

\begin{figure*}
    \includegraphics[width=\textwidth]{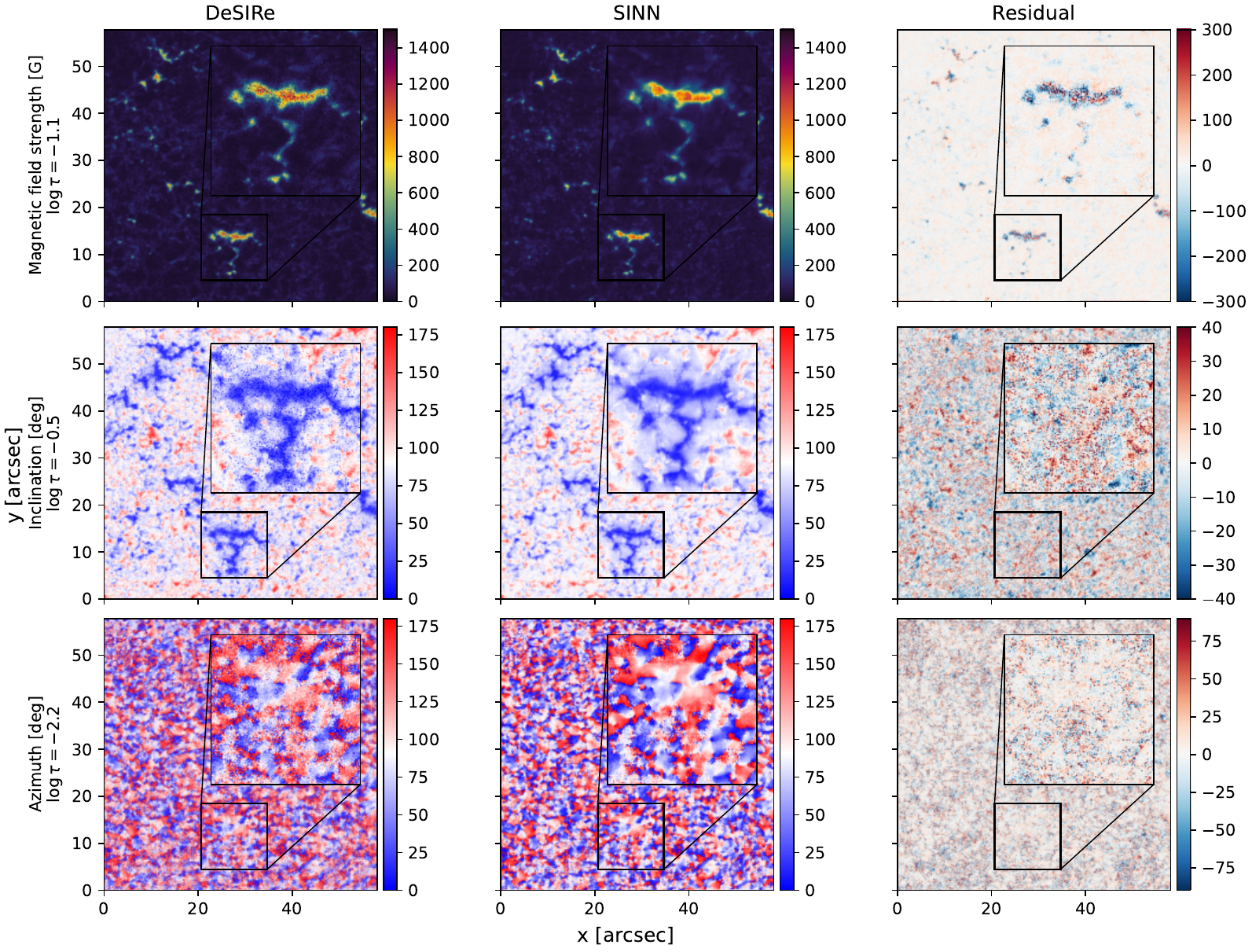}
    \caption{Comparison of the magnetic field strength, inclination, and azimuth inferred from the SCIP observations using \textsc{DeSIRe} (left column) and SINN (middle column). The right column shows the residual difference between the SINN and \textsc{DeSIRe} values; thus positive residuals indicate a larger value predicted by SINN. Insets highlight a network concentration to facilitate comparison of fine-scale magnetic structure. The network element highlighted is analysed in detail by \cite{Campbell2026Network}. An animated figure is available, showing the same figure while stepping through a range of optical depths, $\log\tau = 0.0$ to $-3$. The associated movie is available online.}
    \label{fig:desire_vs_SINN_maps}
\end{figure*}

Figure~\ref{fig:desire_vs_SINN_maps} compares the magnetic field strength, inclination, and azimuth inferred from the SCIP observations using \textsc{DeSIRe} and SINN. An animated version of the figure is available which steps through a range of optical depths, $\log\tau = 0.0$ to $-3$. Once the model had been trained, the SINN/SCIP results presented in this section took only $4.4$ seconds to retrieve on an NVIDIA 5090 GPU. The principal network concentrations identified by the inversion are recovered in the SINN prediction, including the extended magnetic structure highlighted in the inset, which is analysed in detail by \cite{Campbell2026Network}. The spatial morphology of the magnetic field is therefore consistent, indicating that the transformer has learned to identify coherent magnetic structures in the observations. 

The observed differences in the azimuth are largest where there is no detected linear polarisation. DeSIRe most often defaults to $90^\circ$ and SINN defaults to $0^\circ$, but this is of no real physical consequence because the azimuth cannot be constrained without sufficiently informative Stokes $Q$ and $U$ signals. 

The largest absolute differences in magnetic field strength occur in the strongest magnetic elements, but the systematic residual varies as a function of optical depth. At  $\log\tau = -1.1$, the residual is not systematically positive or negative. Deeper in the atmosphere (e.g., $\log\tau = -0.8$), the systematic residual is negative, which indicates that \textsc{DeSIRe} infers magnetic field strengths that are typically larger than those predicted by SINN. Higher in the atmosphere (e.g., $\log\tau = -1.3$), the reverse is the case. Without access to the solar ground truth, we can only speculate as to why this occurs. The lack of a systematic residual at $\log\tau = -1.1$ can be understood as occurring due to the response functions being most sensitive at this layer. The inversion may struggle to accurately stratify  the magnetic field strength near this sensitivity region because \textsc{DeSIRe} places nodes at equidistant locations throughout the full atmosphere and only interpolates between the nodes. In contrast, SINN predicts stratifications that are informed by both self- and cross-attention, and is ultimately trained on simulations that have significant stratifications in all atmospheric parameters. In the network regions, the SINN prediction can be observed to expand as one would expect of a magnetic canopy structure. This behaviour is less clear in the \textsc{DeSIRe} maps. Outside these concentrated flux regions the residuals are comparatively small and largely structureless. 

The SINN inclination map displays less small-scale pixel-to-pixel variability than the \textsc{DeSIRe} inversion while preserving the dominant topology, suggesting that the transformer preferentially recovers spatially coherent magnetic structure. Indeed, there is a significant reduction in so-called inversion noise, which was also the case in \cite{Navarro}. A similar behaviour is observed in the azimuth maps. Although local differences remain, the residuals are largely devoid of extended coherent features, indicating that the principal orientation patterns inferred by \textsc{DeSIRe} are also predicted by SINN. In the animated figure, it should be noted that there is no obvious large-scale residual in the inclination maps in the network element at $\log\tau = -0.5$ and $\log\tau = -3.0$.

\begin{figure*}
    \includegraphics[width=\textwidth]{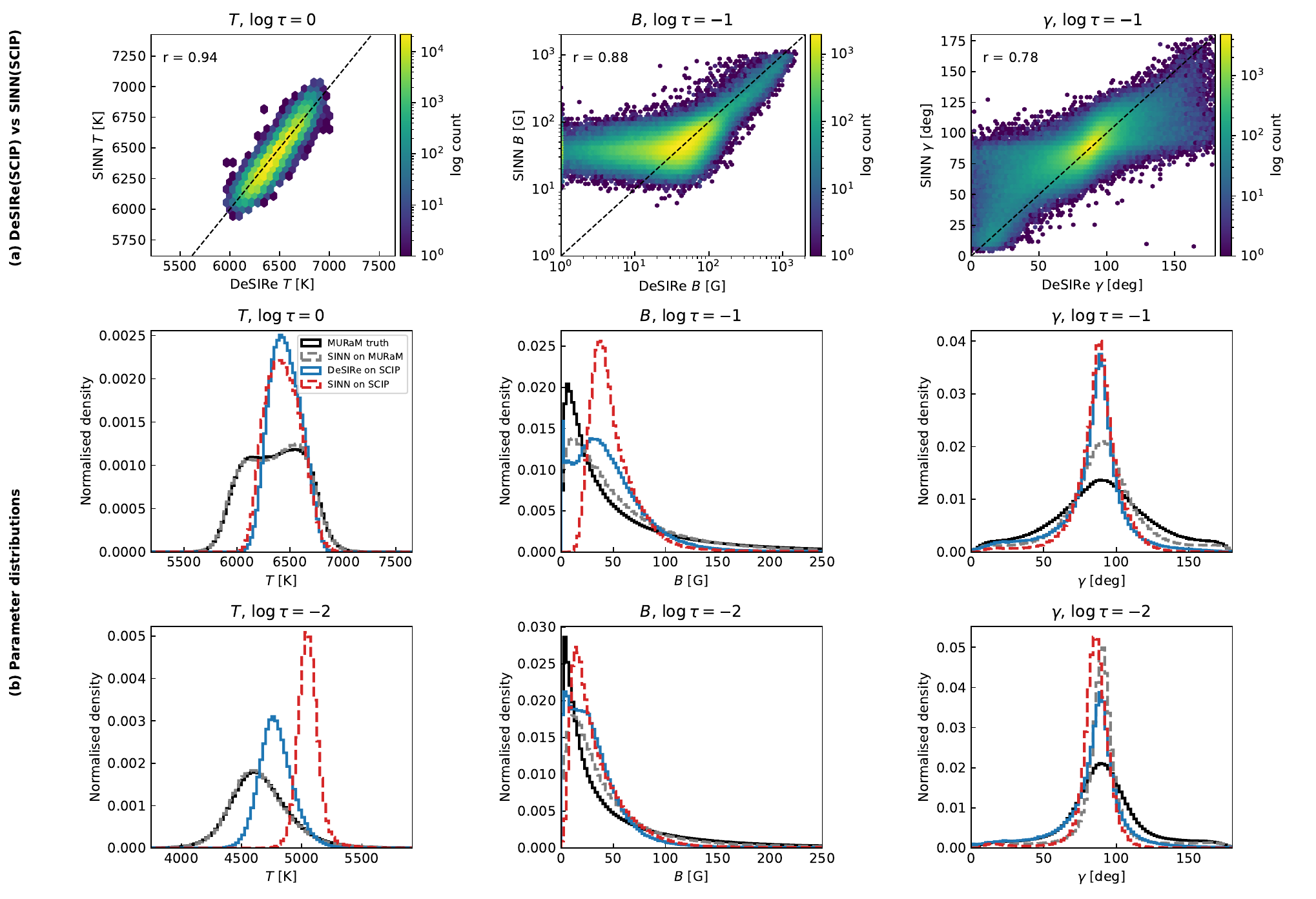}
    \caption{Comparison of atmospheric parameters inferred from the SCIP observations by \textsc{DeSIRe} and SINN at the optical depths used throughout this work. No polarisation-detection threshold is applied. (a) Pixel-by-pixel comparison between the two methods for temperature at $\log\tau=0$, magnetic field strength at $\log\tau=-1$, and magnetic inclination at $\log\tau=-1$. Colours indicate the logarithm of the number of pixels in each bin. The dashed line denotes one-to-one agreement, and the Pearson correlation coefficient is shown in the upper-left corner of each panel. SINN reproduces the principal thermodynamic and magnetic structures inferred by the inversion. (b) Corresponding parameter distributions for the MURaM simulation used for training (black), SINN predictions on the independent MURaM test snapshot (grey), \textsc{DeSIRe} inversions of the SCIP observations (blue), and SINN predictions on the SCIP observations (red dashed). For $B$, only the weakest magnetic field range is shown. }\label{fig:histograms_generalisation2}
\end{figure*}

Figure~\ref{fig:histograms_generalisation2} compares the atmospheric parameters inferred from the SCIP observations by \textsc{DeSIRe} and SINN, and places these results in the context of the MURaM atmospheres used to train the transformer. The upper panels show strong correlations between the two inversion methods for temperature, magnetic field strength, and magnetic inclination. The lower panels compare the corresponding parameter distributions. The distributions inferred by SINN from the SCIP observations differ substantially from those of the MURaM training atmosphere. This demonstrates that, for these observations and parameters, the marginal distributions predicted by SINN are not simply inherited from the marginal atmospheric distributions represented in the training simulation. In Fig.~\ref{fig:histograms_generalisation2}, the comparison between MURaM truth and SINN predictions on the independent MURaM snapshot demonstrates that the transformer generalises to atmospheric structures in an independent snapshot within the same MURaM simulation domain. The corresponding comparison between SINN and \textsc{DeSIRe} on the SCIP observations then assesses whether this learned mapping transfers to real observations. While this comparison cannot establish the absolute \textit{correctness} of either solution in the absence of observational ground truth, it demonstrates that the transformer responds to real solar observations in a physically meaningful manner and converges toward atmospheric statistics consistent with those recovered by a distinct inversion approach. With that said, a discrepancy occurs for the temperature distributions at $\log\tau=-2$, where the spectral line provides substantially weaker constraints. In contrast, the close agreement between the MURaM truth and SINN predictions on the independent test snapshot suggests that this discrepancy is less likely to arise from a failure of the learned atmospheric representation itself. Instead, it points toward limitations imposed by the information content of a single photospheric line and the physical realism of the training atmospheres, both of which may be addressed through future multi-line training on more realistic chromospheric simulations. Another discrepancy between SINN and \textsc{DeSIRe} occurs in the $B$ distributions at $\log\tau=-1$ for the weakest magnetic fields. These distributions include all pixels, including those with very weak or undetectable polarisation, for which the inferred magnetic parameters may be poorly constrained. The closer agreement at $\log\tau=-2$ may partly reflect the greater sensitivity of this spectral line to the horizontal magnetic-field component higher in the photosphere (see Appendix~\ref{sect:appendix_corr}).

Figure~\ref{fig:appendix_smallscale} provides a close-up comparison between the \textsc{DeSIRe} inversion and the SINN (transformer) predictions for a small-scale magnetic feature. Whereas Fig.~\ref{fig:desire_vs_SINN_maps} demonstrates agreement on the scale of the full field of view and in the highlighted network element, this example allows a direct pixel-scale comparison of the inferred magnetic topology across multiple optical depths. The transformer reproduces the principal magnetic structure recovered by \textsc{DeSIRe}, including the spatial organisation of the magnetic field strength, inclination, and azimuth, while producing smoother parameter maps that suppress pixel-to-pixel fluctuations. The agreement is maintained from the deep photosphere to the upper photosphere, indicating that the transformer preserves coherent magnetic morphology not only on large spatial scales but also within weaker, fine-scale magnetic structures. Figure~\ref{fig:appendix_smallscale2} provides a second example, for a magnetic element with more significant linear polarisation than circular polarisation.

\begin{figure*}
    \includegraphics[width=\textwidth]{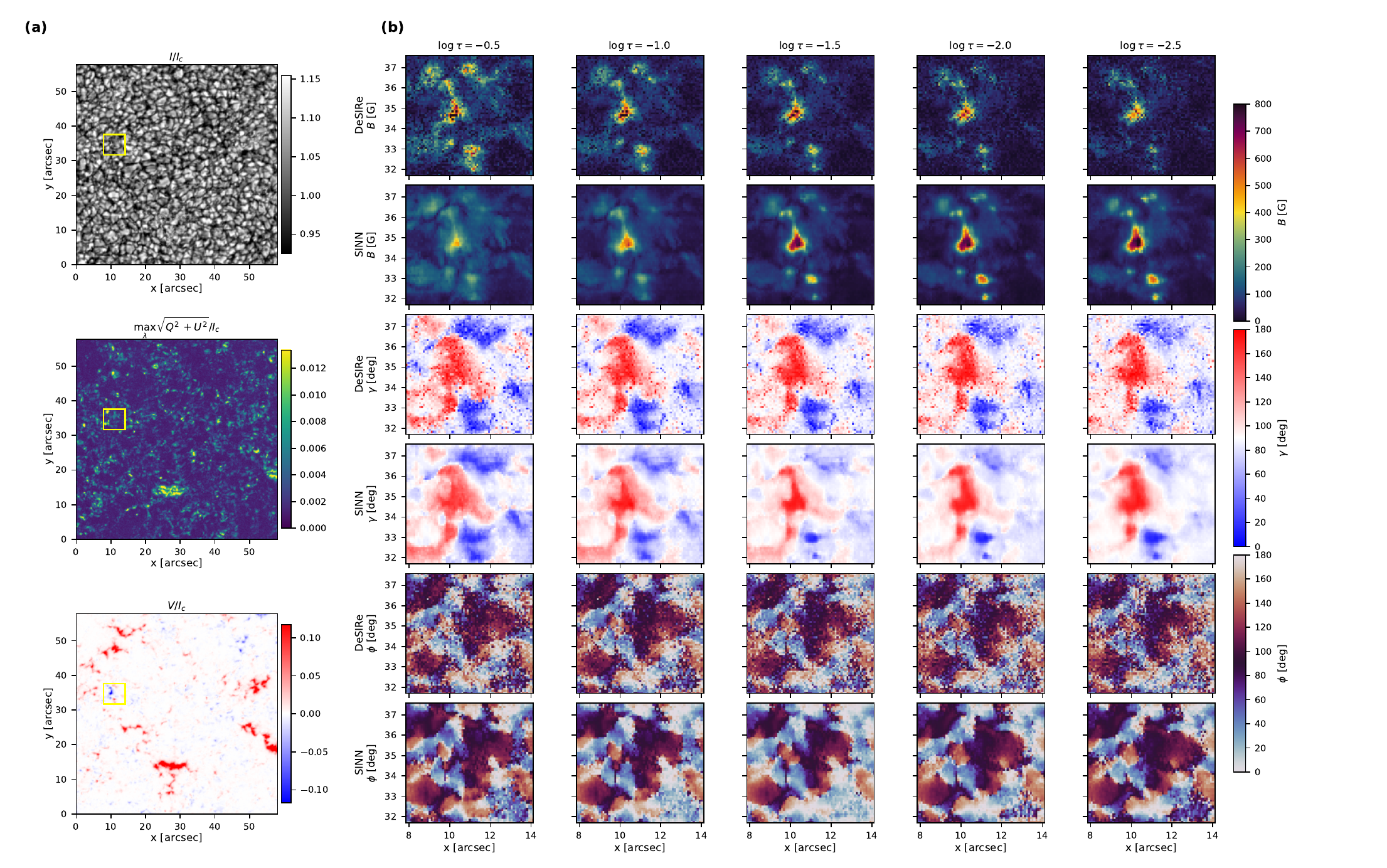}
    \caption{Comparison of \textsc{DeSIRe} and SINN inversions for a small-scale, mixed-polarity magnetic feature. (a) Context maps showing the continuum intensity, maximum linear polarisation, and Stokes $V$ signal, with the selected region indicated by the yellow box. (b) Magnetic field strength, inclination, and azimuth inferred by \textsc{DeSIRe} and SINN at $\log\tau=-0.5$, $-1$, $-1.5$, $-2.0$, and $-2.5$.}\label{fig:appendix_smallscale}
\end{figure*}

\begin{figure*}
    \includegraphics[width=\textwidth]{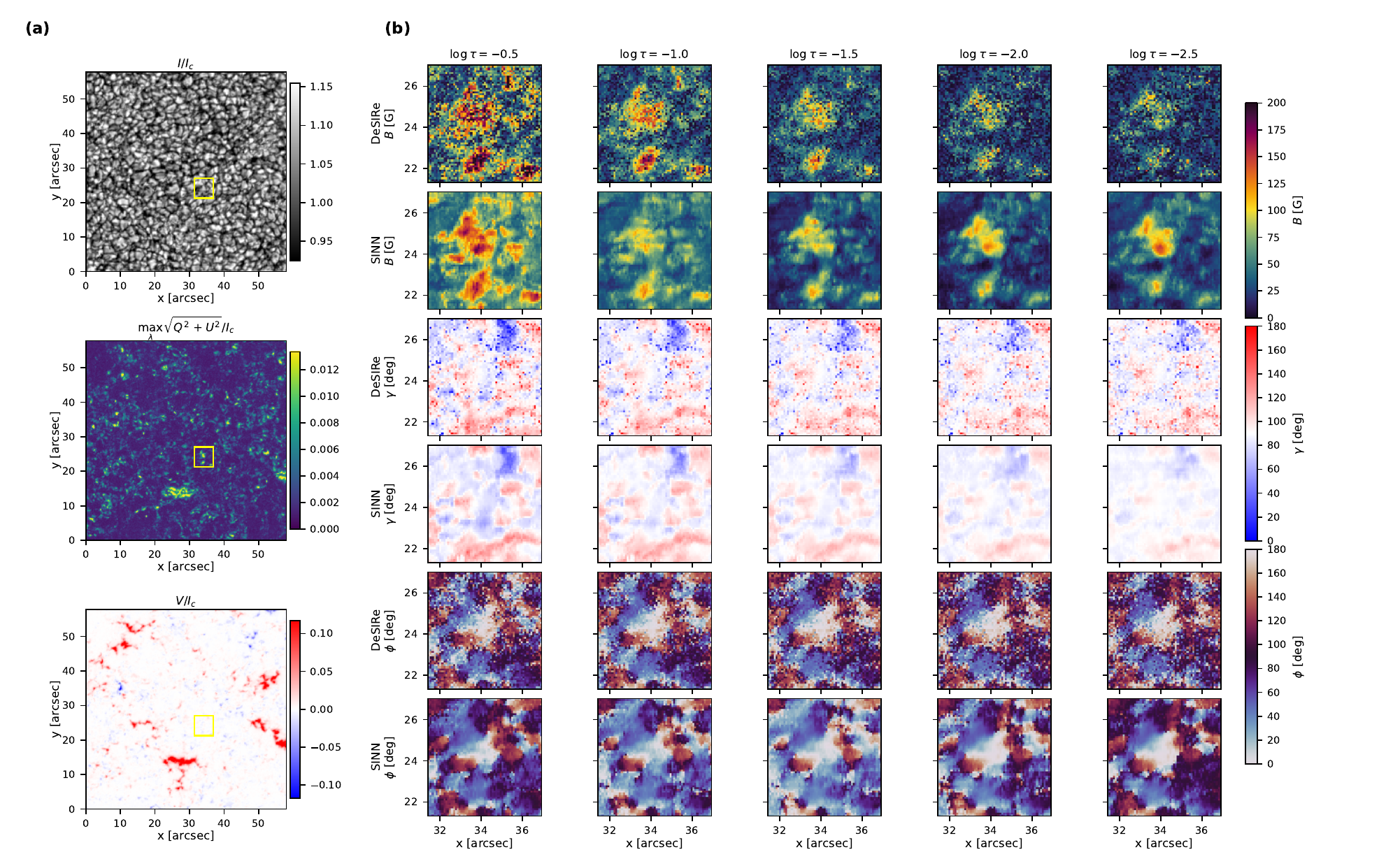}
    \caption{As in Fig.~\ref{fig:appendix_smallscale}, but for a feature with significant linear polarisation and relatively weak circular polarisation. The upper limit on the colourbar for $B$ is significantly lower than in Fig.~\ref{fig:appendix_smallscale}.}\label{fig:appendix_smallscale2}
\end{figure*}

\subsection{Are quiet Sun magnetic fields horizontal or vertical?}\label{sect:ratio}

To investigate how the inferred magnetic topology varies in strongly magnetised and weakly magnetised regions, analysis was first restricted to those exhibiting both significant linear and circular polarisation signals, unlike the all-pixel distributions of total magnetic field strength in Fig.~\ref{fig:histograms_generalisation2}. Specifically, pixels were required to satisfy a $4\sigma$ detection threshold in either Stokes $Q$ or $U$ and in Stokes $V$. This represents about $30\%$ of the pixels. The resulting sample was then divided into two subsets according to the amplitude of Stokes $V$, with pixels above the 75th percentile classified as the strong-signal population and the remainder as the weak-signal population. The classification was performed using the observed Stokes $V$ amplitudes rather than magnetic field strengths inferred from either inversion method to avoid introducing method-dependent selection effects. To facilitate comparison between the two inversion approaches, the magnetic topology was characterised using the median horizontal magnetic field component $B_{\rm h}$, the median unsigned line-of-sight magnetic field component $|B_{\rm LOS}|$, and their ratio,
\begin{equation}
 B_\mathrm{ratio} = \frac{\mathrm{median}~B_{\rm h}}{\mathrm{median}~|B_{\rm LOS}|}.
\end{equation}
To be clear, we first calculate the median $B_{\rm h}$ and median $|B_{\rm LOS}|$ for all pixels considered in each subsample, and then compute one ratio per subsample for every optical depth. We use the median instead of the mean to reduce the influence of outliers in the distributions on our analysis.

\begin{figure*}
    \includegraphics[width=\linewidth]{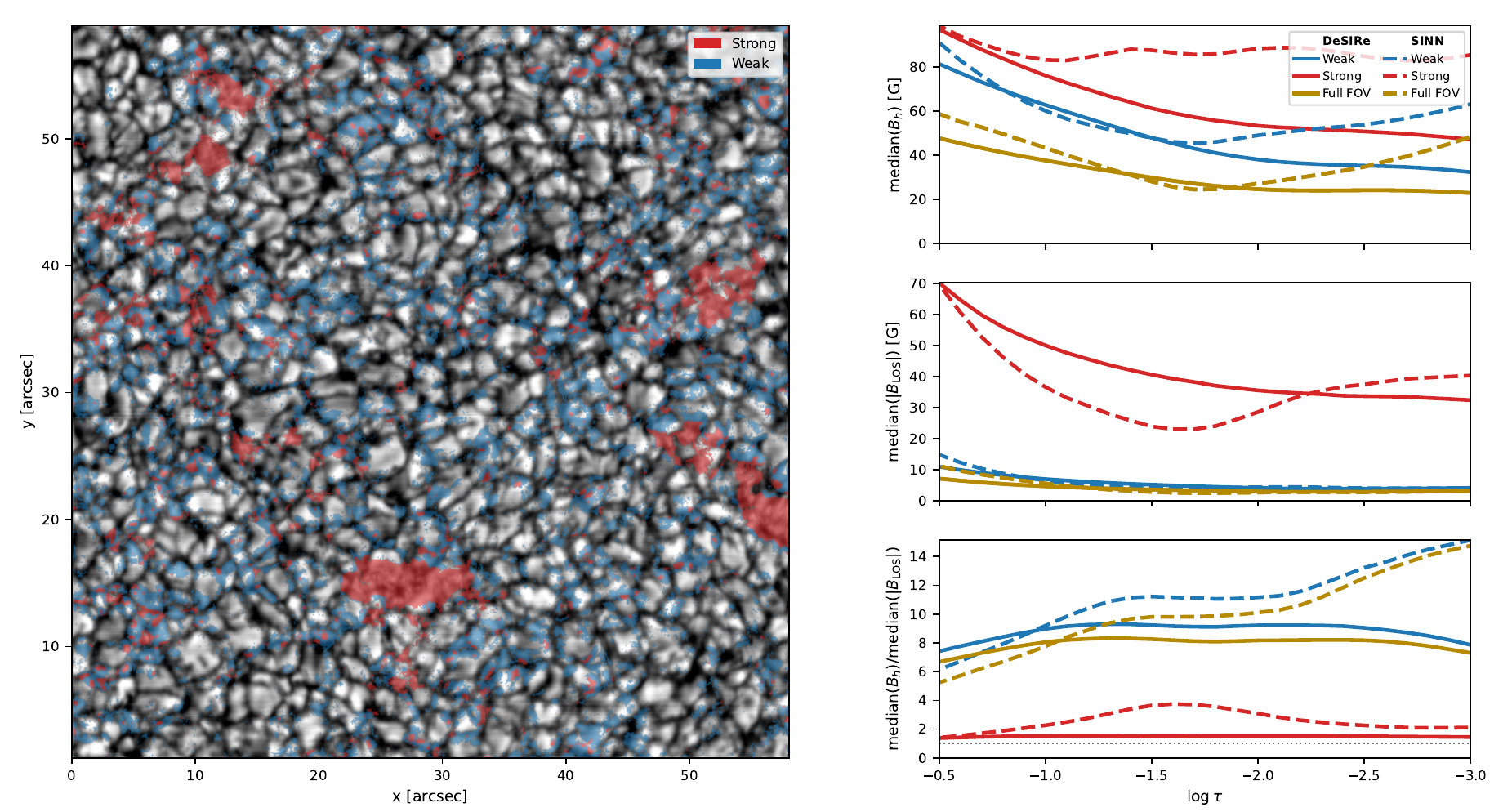}
    \caption{Magnetic topology of weakly and strongly polarised regions inferred from the SCIP observations. The \textit{left panel} shows the selected pixels: either Stokes $Q$ or $U$ and Stokes $V$ must have peak absolute amplitudes exceeding their respective $4\sigma$ noise levels. Among these pixels, the upper quartile of Stokes $V$ amplitude defines the strong-signal population (red), and the remainder defines the weak-signal population (blue). The \textit{right panels} compare the depth dependence of the median horizontal field component $B_h$, the median unsigned line-of-sight component $|B_{\mathrm{LOS}}|$, and their ratio $B_{\mathrm{ratio}}=\mathrm{median}(B_h)/\mathrm{median}(|B_{\mathrm{LOS}}|)$, inferred by \textsc{DeSIRe} (solid) and SINN (dashed). Gold curves show the corresponding statistics for all pixels in the field of view (FOV), i.e. with no polarisation-detection threshold or weak-strong segmentation applied. The dotted line at 1 indicates equal component medians but is not intended to signify an isotropic field.}\label{fig:Bh_vs_Blos}
\end{figure*}

Figure~\ref{fig:Bh_vs_Blos} compares the magnetic topology inferred by \textsc{DeSIRe} and SINN for strong- and weak-field populations identified from the SCIP observations. Both methods recover a clear distinction between the two environments. In the weaker sample, the median horizontal magnetic field component exceeds the median unsigned vertical component throughout the atmosphere, with their $B_\mathrm{ratio}$ reaching values of approximately nine near $\log\tau=-1.0$. In contrast, the stronger population of pixels exhibits a substantially more balanced magnetic configuration, with horizontal and vertical field strengths remaining of a much more comparable magnitude over the photosphere (between $1.5$ and $2.3$ at $\log\tau=-1.0$). These trends are recovered independently by both approaches despite their different methodologies. At $\log\tau=-1.0$, the median horizontal and unsigned line-of-sight field components in the weak-field population is $63$ and $7$~G, respectively, for \textsc{DeSIRe}, compared with $60$ and $6.5$~G for SINN. More detail on these quantitative statistics is available in Appendix~\ref{sect:appendix_B}. A test of the impact of the polarisation amplitude threshold is provided in Appendix~\ref{sect:appendix_C}. As the gold lines in Fig.~\ref{fig:Bh_vs_Blos} shows, the data is horizontally dominated even when all pixels are included in the calculation of $B_\mathrm{ratio}$.

The agreement between SINN and \textsc{DeSIRe} is particularly notable around $\log\tau=-1$, where the Fe~{\sc i}~8468~\AA\ line exhibits strong sensitivity to magnetic field parameters. At these depths, both methods recover very similar field strengths and comparable $B_\mathrm{ratio}$ for both strong and weak populations. Where they differ, however, is in the stratifications in the deep- and upper-photosphere. Most strikingly, the ratio predicted by SINN increases to as high as fourteen at $\log\tau = -3.0$.

Another significant difference occurs in the deepest photospheric layers. Both methods infer stronger magnetic field strengths toward $\log\tau=0$, but SINN predicts a stronger increase than \textsc{DeSIRe}. Consequently, the inferred ratios diverge again in the deep atmosphere, however as the spectral line is less sensitive to these layers this represents an extrapolation with both methods.

To assess whether these differences arise from the inversion configuration rather than the inversion methodology, we repeated the DeSIRe inversions using only the Fe\,{\sc i}\,846.8\,nm line under LTE. Restricting the inversion to the same spectral information used by SINN produced only minor quantitative changes to the inferred magnetic-field stratifications and did not alter the principal conclusions presented in Fig.~\ref{fig:Bh_vs_Blos}. In particular, the upper-photospheric ratio did not increase, such that the discrepancy between \textsc{DeSIRe} and SINN persisted.

Finally, to check whether SINN's predictions are stable with respect to the imposition of a spatial point-spread-function (PSF), we took the synthetic observations generated from the unseen MURaM snapshot and ran SINN on two additional inferences: one where the synthetic profiles were spatially degraded to $0.21\arcsec$ (approximate SCIP resolution), and one to $0.32\arcsec$ (approximate Hinode resolution) by blurring with the appropriate Gaussian 2D PSF. This allows us to compare SINN’s predictions on the native spectra and the degraded spectra. At $\log\tau=-1.2$, the predicted ratio increases from approximately 5.56 at native resolution to $6.92$ and $7.43$ for the degraded cases. The qualitative result persists at both tested PSFs, while the differences between $0.21\arcsec$ and $0.32\arcsec$ are marginal. We also examined the magnitude of the median $B_\mathrm{h}$ and median $|B_\mathrm{LOS}|$ values in each case. We found that the median $B_\mathrm{h}$ is stable, and the changes in $B_\mathrm{ratio}$ are essentially the consequence of relatively small changes in the denominator of $1-2$ G.

\section{Discussion}

Both inversion methodologies recover the same qualitative distinction between strong- and weak-field topology, despite employing different approaches to atmospheric inference. A key question is the extent to which the inferred magnetic topology is consistent with previous studies of quiet-Sun magnetic fields based on Hinode spectropolarimetric observations. Using Hinode observations and inversions, \citet{Suarez2012} found that the internetwork photosphere is characterised by a transverse magnetic field component that is, on average, approximately three times larger than the longitudinal component. \cite{Lites2008} found the apparent transverse magnetic flux density to be five times larger than the longitudinal equivalent. These studies use different methods to arrive at a largely consistent conclusion, but based on comparable short-integration Hinode data that has at best only a few percent of the field-of-view with significant linear polarisation signals in the internetwork quiet Sun \citep{lagg2016, campbell2025_DL}. At $\log\tau=-1.0$, we find using two independent methods and an observational dataset with a much larger fraction of the field-of-view filled with linear polarisation, that the weak-signal (i.e. internetwork) population exhibits $B_\mathrm{ratio}$ of nine. In contrast, the strong-signal population displays systematically smaller values of $B_\mathrm{ratio}$, consistent with stronger magnetic concentrations possessing a relatively larger vertical component. It should be noted, however, that the present analysis is restricted to pixels exhibiting detectable linear and circular polarisation signals. Consequently, this sample is biased toward locations with measurable transverse and longitudinal magnetic fields and does not necessarily represent the full population of quiet-Sun pixels. Nevertheless, that was also true of the analysis presented by \cite{Lites2008,Suarez2012}, and our tests with varying this ratio do not change the conclusions (see Appendix~\ref{sect:appendix_C}). 

The persistence of the discrepancy in the $B_\mathrm{ratio}$ in the upper photosphere between the two inversion methods even when DeSIRe is restricted to the same spectral information as SINN suggests that it is unlikely to arise simply from the inclusion of additional spectral lines in the traditional inversion. Furthermore, the correlation analysis presented in Appendix~\ref{sect:appendix_corr} demonstrates that the Fe\,{\sc i}\,846.8\,nm linear polarisation remains strongly correlated with the horizontal magnetic field component in the mid-to-upper photosphere. The discrepancy therefore cannot be dismissed simply as a consequence of insufficient diagnostic sensitivity of the Fe\,{\sc i}\,846.8\,nm line at these layers. However, the ratio is very sensitive to the denominator and small changes in $\gamma$. Most importantly, given the demonstrated height dependence of the ratio \citep{stenflo2013A&A...555A.132S}, our results should be interpreted both in the context of an  magnetic atmosphere which is expected to become more horizontal in the upper photosphere and in the context of the sensitivity of the line to each component separately (see Appendix~\ref{sect:appendix_corr}).

Despite being trained exclusively on synthetic spectra generated from MURaM, the distributions of atmospheric values inferred by SINN across the {\sc Sunrise~iii} field of view closely match those obtained from independent \textsc{DeSIRe} inversions, at least in the optical depths where the Fe\,{\sc i}\,846.8\,nm line is most sensitive in each parameter. This demonstrates that SINN does not simply reproduce the marginal atmospheric distributions present in its MURaM training set when applied to the SCIP observations. It does not imply that individual predictions are independent of the training data: relationships learned from MURaM may continue to influence or regularise the inferred values, particularly where the observations provide weak constraints. This balance is also depth dependent. A single spectral line does not constrain every atmospheric parameter equally at every optical depth. Consequently, the same observed profile may strongly constrain the magnetic field near the line's principal sensitivity range while leaving deeper or higher layers substantially more dependent on relationships learned from the training atmospheres. The divergence between SINN and \textsc{DeSIRe} in some layers is perhaps consistent with this interpretation. Establishing the relative contributions of observational information and simulation-derived constraints quantitatively will require probabilistic inference and tests across different simulation domains, and represents an important direction for future work.

Some systematic difference between SINN and \textsc{DeSIRe} occurs within the strongest magnetic concentrations, either side of $\log\tau = -1.1$, which is visible in the animated supplementary version of Fig.~\ref{fig:desire_vs_SINN_maps}. One possible explanation for this lies in the limitations of node-based inversions, where interpolation between the nodes can over- or under-estimate the magnetic field strength near the optical depth where peak sensitivity lies. Another possible explanation is that the distribution of Stokes V amplitudes present in the SCIP observations differs from that represented in the synthetic training set. The largest residuals are concentrated in pixels exhibiting strong magnetic signatures, while agreement remains substantially better throughout the weak- and moderate-field population. More generally, this behaviour highlights a broader challenge for supervised inversion methods: differences between the statistical properties of synthetic and observational datasets may lead to systematic biases even when the overall atmospheric structure is recovered successfully. Reducing or mitigating such discrepancies represents an important avenue for future work and will likely be necessary to fully exploit ML inversions on increasingly diverse observational datasets.

Despite the differences discussed above, the overall agreement between SINN and \textsc{DeSIRe} in Figure~\ref{fig:Bh_vs_Blos} is encouraging. The consistency of these trends throughout much of the atmosphere suggests that the principal scientific conclusions are largely independent of the inversion methodology employed. The qualitative stratification of the magnetic topology is largely consistent as predicted by both methods, lending confidence that the transformer captures the principal physical trends present in the observations.

Previous studies have shown that spatially coupled inversions reduce the influence of inversion noise and observational degradation through explicit treatment of the instrumental PSF and spatial regularisation \citep{danilovic2016}. As a simple robustness test, we additionally applied controlled spatial degradation to synthetic full-Stokes spectra prior to SINN inference and found that the qualitative horizontal dominance was preserved under the tested levels of mixing. SINN does not yet employ spatial coupling. Extending transformer-based inversions to incorporate spatial coupling represents a promising avenue for future work. Recent work by \cite{ramos2026arXiv260613004A} indicates this may be beneficial.

\section{Conclusions}

We have presented the first deployment of the transformer-based inversion framework SINN on real spectropolarimetric observations obtained with the SCIP instrument aboard {\sc Sunrise~iii}. Trained exclusively on synthetic spectra generated from radiative magnetohydrodynamic simulations, the model successfully recovered stratified atmospheric parameters from both an independent simulation snapshot and observational data. Comparison with independent \textsc{DeSIRe} inversions demonstrates that SINN reproduces the principal thermodynamic and magnetic structure inferred from the observations, including the distinction between strong and weak magnetic environments and the associated balance between horizontal and vertical magnetic fields.

Perhaps most importantly, our results provide Zeeman-based evidence that the quiet-Sun photosphere contains a substantially larger horizontal magnetic component than inferred from earlier Hinode observations \citep{Lites2008,Suarez2012}. \cite{danilovic2010, danilovic2016} showed that instrumental degradation, inversion methodology, and long temporal integrations can all influence the inferred quiet-Sun inclination distribution. In contrast, the SCIP observations analysed here exhibit abundant linear-polarisation signals without requiring the very long temporal integrations, and the results are obtained using two independent inversion methods. While this does not remove the intrinsic ambiguities of spectropolarimetric inversions, it provides substantially stronger observational constraints on the magnetic field orientation than were available to previous Hinode studies. In this sense, the observations and analyses are consistent with the long-standing picture emerging from Hanle diagnostics that a significant fraction of the quiet-Sun magnetic field has remained inaccessible to previous Zeeman measurements \citep{trujillo2004,tanausu2018ApJ...863..164D,franzi2024ApJ...964...10Z,trelles2021ApJ...915L..20T,franzi2026A&A...708A..97Z}. 

\begin{acknowledgements}
      RJC and MM acknowledge support from the Science
and Technology Facilities Council (STFC) under grant
No.ST/X000923/1. Sunrise III is supported by funding from the Max-Planck-Förderstiftung (Max Planck Foundation), NASA under Grants \#80NSSC18K0934 and \#80NSSC24M0024 (“Heliophysics Low Cost Access to Space” program), and the ISAS/JAXA Small Mission-of-Opportunity program and JSPS KAKENHI Grant Numbers JP18H05234 and JP23K25916. This research has received financial support from the European Union’s Horizon 2020 research and innovation program under grant agreement No. 824135 (SOLARNET) and No. 101097844 (WINSUN) from the European Research Council (ERC). It has also been funded by the Deutsches Zentrum für Luft- und Raumfahrt e.V. (DLR, grant no. 50 OO 1608). The Spanish contributions have been funded by the Spanish MCIN/AEI under projects RTI2018-096886-B-C5, PID2021-125325OB-C5, and PID2024-156066OB-C5, and from “Center of Excellence Severo Ochoa” awards to IAA-CSIC (SEV-2017-0709, CEX2021-001131-S), all co-funded by European REDEF funds, “A way of making Europe".

\end{acknowledgements}

Data availability statement: \textsc{Sunrise III} data are openly available on the Mission Archive (https://sr3data.mps.mpg.de/).

  \bibliographystyle{bibtex/aa} 
  \bibliography{bibtex/biblio}

@ARTICLE{SPIN4D,
       author = {{Yang}, Kai E. and {Tarr}, Lucas A. and {Rempel}, Matthias and {Dodds}, S. Curt and {Jaeggli}, Sarah A. and {Sadowski}, Peter and {Schad}, Thomas A. and {Cunnyngham}, Ian and {Liu}, Jiayi and {Glaser}, Yannik and {Sun}, Xudong},
        title = "{Spectropolarimetric Inversion in Four Dimensions with Deep Learning (SPIn4D). I. Overview, Magnetohydrodynamic Modeling, and Stokes Profile Synthesis}",
      journal = {\apj},
         year = 2024,
        month = dec,
       volume = {976},
       number = {2},
          eid = {204},
        pages = {204},
          doi = {10.3847/1538-4357/ad865b},
archivePrefix = {arXiv},
       eprint = {2407.20309},
 primaryClass = {astro-ph.SR},
       adsurl = {https://ui.adsabs.harvard.edu/abs/2024ApJ...976..204Y}
}

@ARTICLE{rempel,
       author = {{Rempel}, M.},
        title = "{Numerical Sunspot Models: Robustness of Photospheric Velocity and Magnetic Field Structure}",
      journal = {\apj},
         year = 2012,
        month = may,
       volume = {750},
       number = {1},
          eid = {62},
        pages = {62},
          doi = {10.1088/0004-637X/750/1/62},
archivePrefix = {arXiv},
       eprint = {1203.0534},
 primaryClass = {astro-ph.SR},
       adsurl = {https://ui.adsabs.harvard.edu/abs/2012ApJ...750...62R}
}

@ARTICLE{2026arXiv260317929K,
       author = {{Katsukawa}, Y. and {del Toro Iniesta}, J.~C. and {Solanki}, S.~K. and {Kubo}, M. and {Hara}, H. and {Shimizu}, T. and {Oba}, T. and {Kawabata}, Y. and {Tsuzuki}, T. and {Uraguchi}, F. and {Shinoda}, K. and {Tamura}, T. and {Suematsu}, Y. and {Matsumoto}, T. and {Ishikawa}, R.~T. and {Naito}, Y. and {Ichimoto}, K. and {Nagata}, S. and {Anan}, T. and {Orozco Su{\'a}rez}, D. and {Sanchis Kilders}, E. and {Balaguer Jim{\'e}nez}, M. and {L{\'o}pez Jim{\'e}nez}, A.~C. and {Quintero Noda}, C. and {{\'A}lvarez Garc{\'\i}a}, D. and {Ramos M{\'a}s}, J.~L. and {Cobos Carrascosa}, J.~P. and {Labrousse}, P. and {Aparicio del Moral}, B. and {S{\'a}nchez G{\'o}mez}, A. and {Hern{\'a}ndez Exp{\'o}sito}, D. and {Bail{\'o}n Mart{\'\i}nez}, E. and {Morales Fern{\'a}ndez}, J.~M. and {Moreno Mantas}, A. and {Tobaruela}, A. and {Bustamante}, I. and {Bail{\'e}n}, F.~J. and {Blanco Rodr{\'\i}guez}, J. and {Gasent Blesa}, J.~L. and {Rodr{\'\i}guez Mart{\'\i}nez}, P. and {Ferreres}, A. and {Gilabert Palmer}, D. and {Piqueras Carre{\~n}o}, J. and {P{\'e}rez Grande}, I. and {Torralbo}, I. and {{\'A}lvarez-Herrero}, A. and {Korpi-Lagg}, A. and {Gandorfer}, A. and {Berkefeld}, T. and {Bernasconi}, P. and {Feller}, A. and {Riethm{\"u}ller}, T.~L. and {Smitha}, H.~N. and {Mart{\'\i}nez Pillet}, V. and {Grauf}, B. and {Bell}, A. and {Carpenter}, M.},
        title = "{The Sunrise Chromospheric Infrared Spectro-Polarimeter SCIP: An Instrument for SUNRISE III}",
      journal = {\solphys},
         year = 2026,
        month = jul,
       volume = {301},
       number = {7},
          eid = {99},
        pages = {99},
          doi = {10.1007/s11207-026-02696-0},
archivePrefix = {arXiv},
       eprint = {2603.17929},
 primaryClass = {astro-ph.SR},
       adsurl = {https://ui.adsabs.harvard.edu/abs/2026SoPh..301...99K}
}

@ARTICLE{desire,
       author = {{Ruiz Cobo}, B. and {Quintero Noda}, C. and {Gafeira}, R. and {Uitenbroek}, H. and {Orozco Su{\'a}rez}, D. and {P{\'a}ez Ma{\~n}{\'a}}, E.},
        title = "{DeSIRe: Departure coefficient aided Stokes Inversion based on Response functions}",
      journal = {\aap},
         year = 2022,
        month = apr,
       volume = {660},
          eid = {A37},
        pages = {A37},
          doi = {10.1051/0004-6361/202140877},
archivePrefix = {arXiv},
       eprint = {2202.02226},
 primaryClass = {astro-ph.SR},
       adsurl = {https://ui.adsabs.harvard.edu/abs/2022A&A...660A..37R}
}

@ARTICLE{Campbell2026,
       author = {{Campbell}, Ryan J. and {Mathioudakis}, Mihalis and {Quintero Noda}, Carlos},
        title = "{Transformers for Stratified Spectropolarimetric Inversion: Proof of Concept}",
      journal = {\apj},
         year = 2026,
        month = jan,
       volume = {996},
       number = {1},
          eid = {63},
        pages = {63},
          doi = {10.3847/1538-4357/ae1ba5},
archivePrefix = {arXiv},
       eprint = {2506.16810},
 primaryClass = {astro-ph.SR},
       adsurl = {https://ui.adsabs.harvard.edu/abs/2026ApJ...996...63C}
}

@ARTICLE{inversions,
       author = {{del Toro Iniesta}, Jose Carlos and {Ruiz Cobo}, Basilio},
        title = "{Inversion of the radiative transfer equation for polarized light}",
      journal = {Living Reviews in Solar Physics},
         year = 2016,
        month = dec,
       volume = {13},
       number = {1},
          eid = {4},
        pages = {4},
          doi = {10.1007/s41116-016-0005-2},
archivePrefix = {arXiv},
       eprint = {1610.10039},
 primaryClass = {astro-ph.SR},
       adsurl = {https://ui.adsabs.harvard.edu/abs/2016LRSP...13....4D}
}

@ARTICLE{DKIST,
       author = {{Rimmele}, Thomas R. and {Warner}, Mark and {Keil}, Stephen L. and {Goode}, Philip R. and {Kn{\"o}lker}, Michael and {Kuhn}, Jeffrey R. and {Rosner}, Robert R. and {McMullin}, Joseph P. and {Casini}, Roberto and {Lin}, Haosheng and {W{\"o}ger}, Friedrich and {von der L{\"u}he}, Oskar and {Tritschler}, Alexandra and {Davey}, Alisdair and {de Wijn}, Alfred and {Elmore}, David F. and {Fehlmann}, Andr{\'e} and {Harrington}, David M. and {Jaeggli}, Sarah A. and {Rast}, Mark P. and {Schad}, Thomas A. and {Schmidt}, Wolfgang and {Mathioudakis}, Mihalis and {Mickey}, Donald L. and {Anan}, Tetsu and {Beck}, Christian and {Marshall}, Heather K. and {Jeffers}, Paul F. and {Oschmann}, Jacobus M. and {Beard}, Andrew and {Berst}, David C. and {Cowan}, Bruce A. and {Craig}, Simon C. and {Cross}, Eric and {Cummings}, Bryan K. and {Donnelly}, Colleen and {de Vanssay}, Jean-Benoit and {Eigenbrot}, Arthur D. and {Ferayorni}, Andrew and {Foster}, Christopher and {Galapon}, Chriselle Ann and {Gedrites}, Christopher and {Gonzales}, Kerry and {Goodrich}, Bret D. and {Gregory}, Brian S. and {Guzman}, Stephanie S. and {Guzzo}, Stephen and {Hegwer}, Steve and {Hubbard}, Robert P. and {Hubbard}, John R. and {Johansson}, Erik M. and {Johnson}, Luke C. and {Liang}, Chen and {Liang}, Mary and {McQuillen}, Isaac and {Mayer}, Christopher and {Newman}, Karl and {Onodera}, Brialyn and {Phelps}, LeEllen and {Puentes}, Myles M. and {Richards}, Christopher and {Rimmele}, Lukas M. and {Sekulic}, Predrag and {Shimko}, Stephan R. and {Simison}, Brett E. and {Smith}, Brett and {Starman}, Erik and {Sueoka}, Stacey R. and {Summers}, Richard T. and {Szabo}, Aimee and {Szabo}, Louis and {Wampler}, Stephen B. and {Williams}, Timothy R. and {White}, Charles},
        title = "{The Daniel K. Inouye Solar Telescope - Observatory Overview}",
      journal = {\solphys},
         year = 2020,
        month = dec,
       volume = {295},
       number = {12},
          eid = {172},
        pages = {172},
          doi = {10.1007/s11207-020-01736-7},
       adsurl = {https://ui.adsabs.harvard.edu/abs/2020SoPh..295..172R}
}

@ARTICLE{EST,
       author = {{Collados}, M. and {Bettonvil}, F. and {Cavaller}, L. and {Ermolli}, I. and {Gelly}, B. and {P{\'e}rez}, A. and {Socas-Navarro}, H. and {Soltau}, D. and {Volkmer}, R. and {EST Team}},
        title = "{The European Solar Telescope}",
      journal = {\memsai},
         year = 2013,
        month = jan,
       volume = {84},
        pages = {379},
       adsurl = {https://ui.adsabs.harvard.edu/abs/2013MmSAI..84..379C}
}

@ARTICLE{Navarro,
       author = {{Socas-Navarro}, H. and {Asensio Ramos}, A.},
        title = "{Exploring the Sun's upper atmosphere with neural networks: Reversed patterns and the hot wall effect}",
      journal = {\aap},
         year = 2021,
        month = aug,
       volume = {652},
          eid = {A78},
        pages = {A78},
          doi = {10.1051/0004-6361/202140424},
       adsurl = {https://ui.adsabs.harvard.edu/abs/2021A&A...652A..78S}
}

@ARTICLE{Ramos2019,
       author = {{Asensio Ramos}, A. and {D{\'\i}az Baso}, C.~J.},
        title = "{Stokes inversion based on convolutional neural networks}",
      journal = {\aap},
         year = 2019,
        month = jun,
       volume = {626},
          eid = {A102},
        pages = {A102},
          doi = {10.1051/0004-6361/201935628},
archivePrefix = {arXiv},
       eprint = {1904.03714},
 primaryClass = {astro-ph.SR},
       adsurl = {https://ui.adsabs.harvard.edu/abs/2019A&A...626A.102A}
}

@ARTICLE{Ramos2025,
       author = {{D{\'\i}az Baso}, C.~J. and {Asensio Ramos}, A. and {de la Cruz Rodr{\'\i}guez}, J. and {da Silva Santos}, J.~M. and {Rouppe van der Voort}, L.},
        title = "{Exploring spectropolarimetric inversions using neural fields: Solar chromospheric magnetic field under the weak-field approximation}",
      journal = {\aap},
         year = 2025,
        month = jan,
       volume = {693},
          eid = {A170},
        pages = {A170},
          doi = {10.1051/0004-6361/202452172},
archivePrefix = {arXiv},
       eprint = {2409.05156},
 primaryClass = {astro-ph.SR},
       adsurl = {https://ui.adsabs.harvard.edu/abs/2025A&A...693A.170D}
}

@ARTICLE{ramos_jaime_2025,
       author = {{Asensio Ramos}, A. and {de la Cruz Rodr{\'\i}guez}, J.},
        title = "{Neural translation for Stokes inversion and synthesis}",
      journal = {\aap},
         year = 2025,
        month = nov,
       volume = {703},
          eid = {A55},
        pages = {A55},
          doi = {10.1051/0004-6361/202556195},
archivePrefix = {arXiv},
       eprint = {2507.00594},
 primaryClass = {astro-ph.IM},
       adsurl = {https://ui.adsabs.harvard.edu/abs/2025A&A...703A..55A}
}

@ARTICLE{RamosLRSP,
       author = {{Asensio Ramos}, Andr{\'e}s and {Cheung}, Mark C.~M. and {Chifu}, Iulia and {Gafeira}, Ricardo},
        title = "{Machine learning in solar physics}",
      journal = {Living Reviews in Solar Physics},
         year = 2023,
        month = dec,
       volume = {20},
       number = {1},
          eid = {4},
        pages = {4},
          doi = {10.1007/s41116-023-00038-x},
archivePrefix = {arXiv},
       eprint = {2306.15308},
 primaryClass = {astro-ph.SR},
       adsurl = {https://ui.adsabs.harvard.edu/abs/2023LRSP...20....4A}
}

@inproceedings{vaswani2017attention,
  title={Attention is all you need},
  author={Vaswani, Ashish and Shazeer, Noam and Parmar, Niki and Uszkoreit, Jakob and Jones, Llion and Gomez, Aidan N and Kaiser, {\L}ukasz and Polosukhin, Illia},
  booktitle={Advances in Neural Information Processing Systems},
  volume={30},
  year={2017}
}

@article{Campbell2026Network,
       author = {{Campbell}, Ryan J. and {Quintero Noda}, Carlos and {Mathioudakis}, Mihalis and {Collados}, Manuel and {McFall}, Jack and {Jess}, David B. and {Solanki}, Sami K. and {Lagg}, Andreas and {Gandorfer}, Achim and {del Toro Iniesta}, Jose Carlos and {Katsukawa}, Yukio and {Bernasconi}, Pietro and {Berkefeld}, Thomas and {Feller}, Alex and {Riethm{\"u}ller}, Tino L. and {{\'A}lvarez-Herrero}, Alberto and {Kubo}, Masahito and {Smitha}, H.~N. and {Orozco Su{\'a}rez}, David and {Grauf}, Bianca and {Carpenter}, Michael and {Bell}, Alexander and {Mart{\'\i}nez Pillet}, Valent{\'\i}n and {Bail{\'e}n}, Francisco Javier and {Blanco Rodr{\'\i}guez}, Julian and {Castellanos Dur{\'a}n}, Juan Sebasti{\'a}n and {Harnes}, Edvarda and {H{\"o}lken}, Johannes and {Iglesias}, Francisco A. and {Ishikawa}, Ryohtaroh T. and {Kawabata}, Yusuke and {Matsumoto}, Takuma and {Oba}, Takayoshi and {Siu-Tapia}, Azaymi L. and {Strecker}, Hanna and {Vukadinovi{\'c}}, Du{\v{s}}an and {Hara}, Hirohisa and {Shimizu}, Toshifumi},
        title = "{Expanding Magnetic Canopy Structure and Parasitic Polarities in a Quiet-Sun Network Element Observed by SUNRISE III/SCIP}",
      journal = {\apjl},
         year = 2026,
        month = sep,
       volume = {1008},
       number = {1},
          eid = {L19},
        pages = {L19},
          doi = {10.3847/2041-8213/ae946e},
archivePrefix = {arXiv},
       eprint = {2608.03807},
 primaryClass = {astro-ph.SR},
       adsurl = {https://ui.adsabs.harvard.edu/abs/2026ApJ..1008L..19C}
}

@ARTICLE{Lites2017,
       author = {{Lites}, B.~W. and {Rempel}, M. and {Borrero}, J.~M. and {Danilovic}, S.},
        title = "{Are Internetwork Magnetic Fields in the Solar Photosphere Horizontal or Vertical?}",
      journal = {\apj},
         year = 2017,
        month = jan,
       volume = {835},
       number = {1},
          eid = {14},
        pages = {14},
          doi = {10.3847/1538-4357/835/1/14},
       adsurl = {https://ui.adsabs.harvard.edu/abs/2017ApJ...835...14L}
}

@ARTICLE{Lites2008,
       author = {{Lites}, B.~W. and {Kubo}, M. and {Socas-Navarro}, H. and {Berger}, T. and {Frank}, Z. and {Shine}, R. and {Tarbell}, T. and {Title}, A. and {Ichimoto}, K. and {Katsukawa}, Y. and {Tsuneta}, S. and {Suematsu}, Y. and {Shimizu}, T. and {Nagata}, S.},
        title = "{The Horizontal Magnetic Flux of the Quiet-Sun Internetwork as Observed with the Hinode Spectro-Polarimeter}",
      journal = {\apj},
         year = 2008,
        month = jan,
       volume = {672},
       number = {2},
        pages = {1237-1253},
          doi = {10.1086/522922},
       adsurl = {https://ui.adsabs.harvard.edu/abs/2008ApJ...672.1237L}
}

@ARTICLE{Suarez2012,
       author = {{Orozco Su{\'a}rez}, D. and {Bellot Rubio}, L.~R.},
        title = "{Analysis of Quiet-Sun Internetwork Magnetic Fields Based on Linear Polarization Signals}",
      journal = {\apj},
         year = 2012,
        month = may,
       volume = {751},
       number = {1},
          eid = {2},
        pages = {2},
          doi = {10.1088/0004-637X/751/1/2},
archivePrefix = {arXiv},
       eprint = {1203.1440},
 primaryClass = {astro-ph.SR},
       adsurl = {https://ui.adsabs.harvard.edu/abs/2012ApJ...751....2O}
}

@ARTICLE{Campbell2023,
       author = {{Campbell}, Ryan J. and {Gafeira}, Ricardo and {Mathioudakis}, Mihalis and {Noda}, C. Quintero and {Collados}, Manuel},
        title = "{Exploring Magnetic Loops and Serpentine Fields in the Quiet Sun with the GRIS-IFU}",
      journal = {\apj},
         year = 2023,
        month = feb,
       volume = {944},
       number = {2},
          eid = {150},
        pages = {150},
          doi = {10.3847/1538-4357/acb33e},
archivePrefix = {arXiv},
       eprint = {2301.05591},
 primaryClass = {astro-ph.SR},
       adsurl = {https://ui.adsabs.harvard.edu/abs/2023ApJ...944..150C}
}

@ARTICLE{sunrise3,
       author = {{Korpi-Lagg}, Andreas and {Gandorfer}, Achim and {Solanki}, Sami K. and {del Toro Iniesta}, Jose Carlos and {Katsukawa}, Yukio and {Bernasconi}, Pietro and {Berkefeld}, Thomas and {Feller}, Alex and {Riethm{\"u}ller}, Tino L. and {{\'A}lvarez-Herrero}, Alberto and {Kubo}, Masahito and {Mart{\'\i}nez Pillet}, Valent{\'\i}n and {Smitha}, H.~N. and {Orozco Su{\'a}rez}, David and {Grauf}, Bianca and {Carpenter}, Michael and {Bell}, Alexander and {{\'A}lvarez-Alonso}, Mar{\'\i}a-Teresa and {{\'A}lvarez Garc{\'\i}a}, Daniel and {Aparicio del Moral}, Beatriz and {Ati{\'e}nzar}, Julia and {Ayoub}, Daniel and {Bail{\'e}n}, Francisco Javier and {Bail{\'o}n Mart{\'\i}nez}, Eduardo and {Balaguer Jim{\'e}nez}, Maria and {Barthol}, Peter and {Bayon Laguna}, Montserrat and {Bellot Rubio}, Luis R. and {Bergmann}, Melani and {Blanco Rodr{\'\i}guez}, Julian and {Bochmann}, Jan and {Borrero}, Juan Manuel and {Campos-Jara}, Antonio and {Castellanos Dur{\'a}n}, Juan Sebasti{\'a}n and {Cebollero}, Mar{\'\i}a and {Conde Rodr{\'\i}guez}, Aitor and {Deutsch}, Werner and {Eaton}, Harry and {Fern{\'a}ndez-Medina}, Ana Belen and {Fernandez-Rico}, German and {Ferreres}, Agustin and {Garc{\'\i}a}, Andr{\'e}s and {Garc{\'\i}a Alarcia}, Ram{\'o}n Mar{\'\i}a and {Garc{\'\i}a Parejo}, Pilar and {Garranzo-Garc{\'\i}a}, Daniel and {Gasent Blesa}, Jos{\'e} Luis and {Gerber}, Karin and {Germerott}, Dietmar and {Gilabert Palmer}, David and {Gizon}, Laurent and {G{\'o}mez S{\'a}nchez-Tirado}, Miguel Angel and {Gonz{\'a}lez-B{\'a}rcena}, David and {Gonzalo Melchor}, Alejandro and {Goodyear}, Sam and {Hara}, Hirohisa and {Harnes}, Edvarda and {Heerlein}, Klaus and {Heidecke}, Frank and {Heinrichs}, Jan and {Hern{\'a}ndez Exp{\'o}sito}, David and {Hirzberger}, Johann and {Hoelken}, Johannes and {Hyun}, Sangwon and {Iglesias}, Francisco A. and {Ishikawa}, Ryohtaroh T. and {Jeon}, Minwoo and {Kawabata}, Yusuke and {Kolleck}, Martin and {Laguna}, Hugo and {Lomas}, Julian and {L{\'o}pez Jim{\'e}nez}, Antonio C. and {Manzano}, Paula and {Matsumoto}, Takuma and {Mayo Turrado}, David and {Meierdierks}, Thimo and {Meining}, Stefan and {Monecke}, Markus and {Morales-Fern{\'a}ndez}, Jos{\'e} Miguel and {Moreno Mantas}, Antonio Jes{\'u}s and {Moreno Vacas}, Alejandro and {M{\"u}ller}, Marc Ferenc and {M{\"u}ller}, Reinhard and {Naito}, Yoshihiro and {Nakai}, Eiji and {N{\'u}{\~n}ez Peral}, Armon{\'\i}a and {Oba}, Takayoshi and {Palo}, Geoffrey and {P{\'e}rez-Grande}, Isabel and {Piqueras Carre{\~n}o}, Javier and {Preis}, Tobias and {Przybylski}, Damien and {Quintero Noda}, Carlos and {Ramanath}, Sandeep and {Ramos M{\'a}s}, Jose Luis and {Raouafi}, Nour and {Rivas-Mart{\'\i}nez}, Mar{\'\i}a-Jes{\'u}s and {Rodr{\'\i}guez Mart{\'\i}nez}, Pedro and {Rodr{\'\i}guez Valido}, Manuel and {Ruiz Cobo}, Basilio and {S{\'a}nchez Rodr{\'\i}guez}, Antonio and {Sanchez Toledo}, Mariano and {S{\'a}nchez G{\'o}mez}, Antonio and {Sanchis Kilders}, Esteban and {Sant}, Kamal and {Santamarina Guerrero}, Pablo and {Schulze}, Erich and {Shimizu}, Toshifumi and {Silva-L{\'o}pez}, Manuel and {Singh}, Kunal and {Siu-Tapia}, Azaymi L. and {Sonner}, Thomas and {Staub}, Jan and {Strecker}, Hanna and {Tobaruela}, Angel and {Torralbo}, Ignacio and {Tritschler}, Alexandra and {Tsuzuki}, Toshihiro and {Uraguchi}, Fumihiro and {Volkmer}, Reiner and {Vourlidas}, Angelos and {Vukadinovi{\'c}}, Du{\v{s}}an and {Werner}, Stephan and {Zerr}, Andreas},
        title = "{SUNRISE III: Overview of Observatory and Instruments}",
      journal = {\solphys},
         year = 2025,
        month = may,
       volume = {300},
       number = {5},
          eid = {75},
        pages = {75},
          doi = {10.1007/s11207-025-02485-1},
archivePrefix = {arXiv},
       eprint = {2502.06483},
 primaryClass = {astro-ph.IM},
       adsurl = {https://ui.adsabs.harvard.edu/abs/2025SoPh..300...75K}
}

@ARTICLE{lagg2016,
       author = {{Lagg}, A. and {Solanki}, S.~K. and {Doerr}, H.-P. and {Mart{\'\i}nez Gonz{\'a}lez}, M.~J. and {Riethm{\"u}ller}, T. and {Collados Vera}, M. and {Schlichenmaier}, R. and {Orozco Su{\'a}rez}, D. and {Franz}, M. and {Feller}, A. and {Kuckein}, C. and {Schmidt}, W. and {Asensio Ramos}, A. and {Pastor Yabar}, A. and {von der L{\"u}he}, O. and {Denker}, C. and {Balthasar}, H. and {Volkmer}, R. and {Staude}, J. and {Hofmann}, A. and {Strassmeier}, K. and {Kneer}, F. and {Waldmann}, T. and {Borrero}, J.~M. and {Sobotka}, M. and {Verma}, M. and {Louis}, R.~E. and {Rezaei}, R. and {Soltau}, D. and {Berkefeld}, T. and {Sigwarth}, M. and {Schmidt}, D. and {Kiess}, C. and {Nicklas}, H.},
        title = "{Probing deep photospheric layers of the quiet Sun with high magnetic sensitivity}",
      journal = {\aap},
         year = 2016,
        month = nov,
       volume = {596},
          eid = {A6},
        pages = {A6},
          doi = {10.1051/0004-6361/201628489},
archivePrefix = {arXiv},
       eprint = {1605.06324},
 primaryClass = {astro-ph.SR},
       adsurl = {https://ui.adsabs.harvard.edu/abs/2016A&A...596A...6L}
}

@ARTICLE{campbell2025_DL,
       author = {{Campbell}, Ryan J. and {Mathioudakis}, M. and {Quintero Noda}, Carlos and {Keys}, P.~H. and {Orozco Su{\'a}rez}, D.},
        title = "{Application of Deep Learning to the Classification of Stokes Profiles: From the Quiet Sun to Sunspots}",
      journal = {\apj},
         year = 2025,
        month = jul,
       volume = {988},
       number = {1},
          eid = {9},
        pages = {9},
          doi = {10.3847/1538-4357/addb49},
archivePrefix = {arXiv},
       eprint = {2505.14275},
 primaryClass = {astro-ph.SR},
       adsurl = {https://ui.adsabs.harvard.edu/abs/2025ApJ...988....9C}
}

@ARTICLE{trujillo2004,
       author = {{Trujillo Bueno}, J. and {Shchukina}, N. and {Asensio Ramos}, A.},
        title = "{A substantial amount of hidden magnetic energy in the quiet Sun}",
      journal = {\nat},
         year = 2004,
        month = jul,
       volume = {430},
       number = {6997},
        pages = {326-329},
          doi = {10.1038/nature02669},
archivePrefix = {arXiv},
       eprint = {astro-ph/0409004},
 primaryClass = {astro-ph},
       adsurl = {https://ui.adsabs.harvard.edu/abs/2004Natur.430..326T}
}

@ARTICLE{tanausu2018ApJ...863..164D,
       author = {{del Pino Alem{\'a}n}, T. and {Trujillo Bueno}, J. and {{\v{S}}t{\v{e}}p{\'a}n}, J. and {Shchukina}, N.},
        title = "{A Novel Investigation of the Small-scale Magnetic Activity of the Quiet Sun via the Hanle Effect in the Sr I 4607 {\r{A}} Line}",
      journal = {\apj},
         year = 2018,
        month = aug,
       volume = {863},
       number = {2},
          eid = {164},
        pages = {164},
          doi = {10.3847/1538-4357/aaceab},
archivePrefix = {arXiv},
       eprint = {1806.07293},
 primaryClass = {astro-ph.SR},
       adsurl = {https://ui.adsabs.harvard.edu/abs/2018ApJ...863..164D}
}

@ARTICLE{franzi2024ApJ...964...10Z,
       author = {{Zeuner}, Franziska and {del Pino Alem{\'a}n}, Tanaus{\'u} and {Trujillo Bueno}, Javier and {Solanki}, Sami K.},
        title = "{Comparing Observed with Simulated Solar-disk-center Scattering Polarization in the Sr I 4607 {\r{A}} Line}",
      journal = {\apj},
         year = 2024,
        month = mar,
       volume = {964},
       number = {1},
          eid = {10},
        pages = {10},
          doi = {10.3847/1538-4357/ad26f9},
archivePrefix = {arXiv},
       eprint = {2402.04736},
 primaryClass = {astro-ph.SR},
       adsurl = {https://ui.adsabs.harvard.edu/abs/2024ApJ...964...10Z}
}

@ARTICLE{franzi2026A&A...708A..97Z,
       author = {{Zeuner}, Franziska and {Alsina Ballester}, Ernest and {Belluzzi}, Luca and {Casini}, Roberto and {Harrington}, David M. and {del Pino Alem{\'a}n}, Tanaus{\'u} and {Trujillo Bueno}, Javier},
        title = "{Observational insights into Sr I 4607 {\r{A}} scattering polarization with DKIST/ViSP}",
      journal = {\aap},
         year = 2026,
        month = apr,
       volume = {708},
          eid = {A97},
        pages = {A97},
          doi = {10.1051/0004-6361/202556974},
archivePrefix = {arXiv},
       eprint = {2603.01854},
 primaryClass = {astro-ph.SR},
       adsurl = {https://ui.adsabs.harvard.edu/abs/2026A&A...708A..97Z}
}

@ARTICLE{solanki2026,
       author = {{Solanki}, Sami K. and {Smitha}, H.~N. and {Lagg}, Andreas and {Gandorfer}, Achim and {del Toro Iniesta}, Jose Carlos and {Katsukawa}, Yukio and {Bernasconi}, Pietro and {Berkefeld}, Thomas and {Feller}, Alex and {Riethm{\"u}ller}, Tino L. and et al.},
        title = "{SUNRISE III: Instrument, Mission, Data, and First Results}",
      journal = {\apjl},
         year = 2026,
        month = jul,
       volume = {1005},
       number = {2},
          eid = {L64},
        pages = {L64},
          doi = {10.3847/2041-8213/ae796b},
archivePrefix = {arXiv},
       eprint = {2606.07989},
 primaryClass = {astro-ph.SR},
       adsurl = {https://ui.adsabs.harvard.edu/abs/2026ApJ..1005L..64S}
}

@ARTICLE{danilovic2016,
       author = {{Danilovic}, S. and {van Noort}, M. and {Rempel}, M.},
        title = "{Internetwork magnetic field as revealed by two-dimensional inversions}",
      journal = {\aap},
         year = 2016,
        month = sep,
       volume = {593},
          eid = {A93},
        pages = {A93},
          doi = {10.1051/0004-6361/201527842},
archivePrefix = {arXiv},
       eprint = {1607.00772},
 primaryClass = {astro-ph.SR},
       adsurl = {https://ui.adsabs.harvard.edu/abs/2016A&A...593A..93D}
}

@ARTICLE{andres2014,
       author = {{Asensio Ramos}, A. and {Mart{\'\i}nez Gonz{\'a}lez}, M.~J.},
        title = "{Hierarchical analysis of the quiet-Sun magnetism}",
      journal = {\aap},
         year = 2014,
        month = dec,
       volume = {572},
          eid = {A98},
        pages = {A98},
          doi = {10.1051/0004-6361/201423860},
archivePrefix = {arXiv},
       eprint = {1410.5953},
 primaryClass = {astro-ph.SR},
       adsurl = {https://ui.adsabs.harvard.edu/abs/2014A&A...572A..98A}
}

@ARTICLE{andres2009,
       author = {{Asensio Ramos}, A.},
        title = "{Evidence for Quasi-Isotropic Magnetic Fields from Hinode Quiet-Sun Observations}",
      journal = {\apj},
         year = 2009,
        month = aug,
       volume = {701},
       number = {2},
        pages = {1032-1043},
          doi = {10.1088/0004-637X/701/2/1032},
archivePrefix = {arXiv},
       eprint = {0906.4230},
 primaryClass = {astro-ph.SR},
       adsurl = {https://ui.adsabs.harvard.edu/abs/2009ApJ...701.1032A}
}

@ARTICLE{borrero2012,
       author = {{Borrero}, J.~M. and {Kobel}, P.},
        title = "{Inferring the magnetic field vector in the quiet Sun. II. Interpreting results from the inversion of Stokes profiles}",
      journal = {\aap},
         year = 2012,
        month = nov,
       volume = {547},
          eid = {A89},
        pages = {A89},
          doi = {10.1051/0004-6361/201118238},
archivePrefix = {arXiv},
       eprint = {1209.4830},
 primaryClass = {astro-ph.SR},
       adsurl = {https://ui.adsabs.harvard.edu/abs/2012A&A...547A..89B}
}

@ARTICLE{borrero2011,
       author = {{Borrero}, J.~M. and {Kobel}, P.},
        title = "{Inferring the magnetic field vector in the quiet Sun. I. Photon noise and selection criteria}",
      journal = {\aap},
         year = 2011,
        month = mar,
       volume = {527},
          eid = {A29},
        pages = {A29},
          doi = {10.1051/0004-6361/201015634},
archivePrefix = {arXiv},
       eprint = {1011.4380},
 primaryClass = {astro-ph.SR},
       adsurl = {https://ui.adsabs.harvard.edu/abs/2011A&A...527A..29B}
}

@ARTICLE{2019LRSP...16....1B,
       author = {{Bellot Rubio}, Luis and {Orozco Su{\'a}rez}, David},
        title = "{Quiet Sun magnetic fields: an observational view}",
      journal = {Living Reviews in Solar Physics},
         year = 2019,
        month = dec,
       volume = {16},
       number = {1},
          eid = {1},
        pages = {1},
          doi = {10.1007/s41116-018-0017-1},
       adsurl = {https://ui.adsabs.harvard.edu/abs/2019LRSP...16....1B}
}

@ARTICLE{kosugi2007SoPh..243....3K,
       author = {{Kosugi}, T. and {Matsuzaki}, K. and {Sakao}, T. and {Shimizu}, T. and {Sone}, Y. and {Tachikawa}, S. and {Hashimoto}, T. and {Minesugi}, K. and {Ohnishi}, A. and {Yamada}, T. and {Tsuneta}, S. and {Hara}, H. and {Ichimoto}, K. and {Suematsu}, Y. and {Shimojo}, M. and {Watanabe}, T. and {Shimada}, S. and {Davis}, J.~M. and {Hill}, L.~D. and {Owens}, J.~K. and {Title}, A.~M. and {Culhane}, J.~L. and {Harra}, L.~K. and {Doschek}, G.~A. and {Golub}, L.},
        title = "{The Hinode (Solar-B) Mission: An Overview}",
      journal = {\solphys},
         year = 2007,
        month = jun,
       volume = {243},
       number = {1},
        pages = {3-17},
          doi = {10.1007/s11207-007-9014-6},
       adsurl = {https://ui.adsabs.harvard.edu/abs/2007SoPh..243....3K}
}

@ARTICLE{lites2013SoPh..283..579L,
       author = {{Lites}, B.~W. and {Akin}, D.~L. and {Card}, G. and {Cruz}, T. and {Duncan}, D.~W. and {Edwards}, C.~G. and {Elmore}, D.~F. and {Hoffmann}, C. and {Katsukawa}, Y. and {Katz}, N. and {Kubo}, M. and {Ichimoto}, K. and {Shimizu}, T. and {Shine}, R.~A. and {Streander}, K.~V. and {Suematsu}, A. and {Tarbell}, T.~D. and {Title}, A.~M. and {Tsuneta}, S.},
        title = "{The Hinode Spectro-Polarimeter}",
      journal = {\solphys},
         year = 2013,
        month = apr,
       volume = {283},
       number = {2},
        pages = {579-599},
          doi = {10.1007/s11207-012-0206-3},
       adsurl = {https://ui.adsabs.harvard.edu/abs/2013SoPh..283..579L}
}

@ARTICLE{campbell2021A&A...647A.182C,
       author = {{Campbell}, R.~J. and {Mathioudakis}, M. and {Collados}, M. and {Keys}, P.~H. and {Asensio Ramos}, A. and {Nelson}, C.~J. and {Kuridze}, D. and {Reid}, A.},
        title = "{Temporal evolution of small-scale internetwork magnetic fields in the solar photosphere}",
      journal = {\aap},
         year = 2021,
        month = mar,
       volume = {647},
          eid = {A182},
        pages = {A182},
          doi = {10.1051/0004-6361/202040028},
archivePrefix = {arXiv},
       eprint = {2102.00942},
 primaryClass = {astro-ph.SR},
       adsurl = {https://ui.adsabs.harvard.edu/abs/2021A&A...647A.182C}
}

@ARTICLE{trelles2021ApJ...915L..20T,
       author = {{Trelles Arjona}, J.~C. and {Mart{\'\i}nez Gonz{\'a}lez}, M.~J. and {Ruiz Cobo}, B.},
        title = "{Mapping the Hidden Magnetic Field of the Quiet Sun}",
      journal = {\apjl},
         year = 2021,
        month = jul,
       volume = {915},
       number = {1},
          eid = {L20},
        pages = {L20},
          doi = {10.3847/2041-8213/ac0af2},
archivePrefix = {arXiv},
       eprint = {2106.10546},
 primaryClass = {astro-ph.SR},
       adsurl = {https://ui.adsabs.harvard.edu/abs/2021ApJ...915L..20T}
}

@ARTICLE{kianfar2018SoPh..293..123K,
       author = {{Kianfar}, S. and {Jafarzadeh}, S. and {Mirtorabi}, M.~T. and {Riethm{\"u}ller}, T.~L.},
        title = "{Linear Polarization Features in the Quiet-Sun Photosphere: Structure and Dynamics}",
      journal = {\solphys},
         year = 2018,
        month = aug,
       volume = {293},
       number = {8},
          eid = {123},
        pages = {123},
          doi = {10.1007/s11207-018-1341-2},
archivePrefix = {arXiv},
       eprint = {1807.04633},
 primaryClass = {astro-ph.SR},
       adsurl = {https://ui.adsabs.harvard.edu/abs/2018SoPh..293..123K}
}

@ARTICLE{stenflo2013A&A...555A.132S,
       author = {{Stenflo}, J.~O.},
        title = "{Horizontal or vertical magnetic fields on the quiet Sun. Angular distributions and their height variations}",
      journal = {\aap},
         year = 2013,
        month = jul,
       volume = {555},
          eid = {A132},
        pages = {A132},
          doi = {10.1051/0004-6361/201321608},
archivePrefix = {arXiv},
       eprint = {1306.3581},
 primaryClass = {astro-ph.SR},
       adsurl = {https://ui.adsabs.harvard.edu/abs/2013A&A...555A.132S}
}

@ARTICLE{yang2025ApJ...995..146Y,
       author = {{Yang}, Kai E. and {Sun}, Xudong and {Tarr}, Lucas A. and {Liu}, Jiayi and {Sadowski}, Peter and {Dodds}, S. Curt and {Rempel}, Matthias and {Jaeggli}, Sarah A. and {Schad}, Thomas A. and {Cunnyngham}, Ian and {Glaser}, Yannik and {Wolniewicz}, Linnea},
        title = "{Spectropolarimetric Inversion in Four Dimensions with Deep Learning (SPIn4D). II. A Physics-informed Machine Learning Method for 3D Solar Photosphere Reconstruction}",
      journal = {\apj},
         year = 2025,
        month = dec,
       volume = {995},
       number = {2},
          eid = {146},
        pages = {146},
          doi = {10.3847/1538-4357/ae12ef},
archivePrefix = {arXiv},
       eprint = {2510.09967},
 primaryClass = {astro-ph.SR},
       adsurl = {https://ui.adsabs.harvard.edu/abs/2025ApJ...995..146Y}
}

@ARTICLE{ramos2026arXiv260613004A,
       author = {{Asensio Ramos}, A. and {Yang}, K.~E. and {Mart{\'\i}nez Gonz{\'a}lez}, M.~J. and {Dodds}, S. Curt and {Sun}, X.},
        title = "{3DStokesFlow: Simulation-based inference for 3D Stokes profiles using flow matching}",
      journal = {\aap},
         year = 2026,
        month = sep,
       volume = {713},
          eid = {A89},
        pages = {A89},
          doi = {10.1051/0004-6361/202661300},
archivePrefix = {arXiv},
       eprint = {2606.13004},
 primaryClass = {astro-ph.SR},
       adsurl = {https://ui.adsabs.harvard.edu/abs/2026A&A...713A..89A}
}

@ARTICLE{danilovic2010,
       author = {{Danilovic}, S. and {Sch{\"u}ssler}, M. and {Solanki}, S.~K.},
        title = "{Probing quiet Sun magnetism using MURaM simulations and Hinode/SP results: support for a local dynamo}",
      journal = {\aap},
         year = 2010,
        month = apr,
       volume = {513},
          eid = {A1},
        pages = {A1},
          doi = {10.1051/0004-6361/200913379},
archivePrefix = {arXiv},
       eprint = {1001.2183},
 primaryClass = {astro-ph.SR},
       adsurl = {https://ui.adsabs.harvard.edu/abs/2010A&A...513A...1D}
}

\begin{appendix}

\section{Correlations between synthesised polarisation signals and magnetic-field components in MURaM}\label{sect:appendix_corr}

To provide additional context for the magnetic-field component ratios shown in Fig.~\ref{fig:Bh_vs_Blos}, we investigated the empirical relationship between the synthesised polarisation signals and the underlying magnetic field in the MURaM simulations used throughout this work. For each spectral region, we computed the Pearson correlation coefficient between the wavelength-integrated linear polarisation,

\begin{equation}
L = \int \sqrt{Q^2(\lambda)+U^2(\lambda)}\,{\rm d}\lambda,
\end{equation}

and the horizontal magnetic field component, $B_h$, as a function of optical depth. An analogous analysis was performed for the wavelength-integrated circular polarisation,

\begin{equation}
V_{\rm int} = \int |V(\lambda)|\,{\rm d}\lambda,
\end{equation}

and the line-of-sight magnetic field, $|B_{\rm LOS}|$.

\begin{figure*}
    \includegraphics[width=\textwidth]{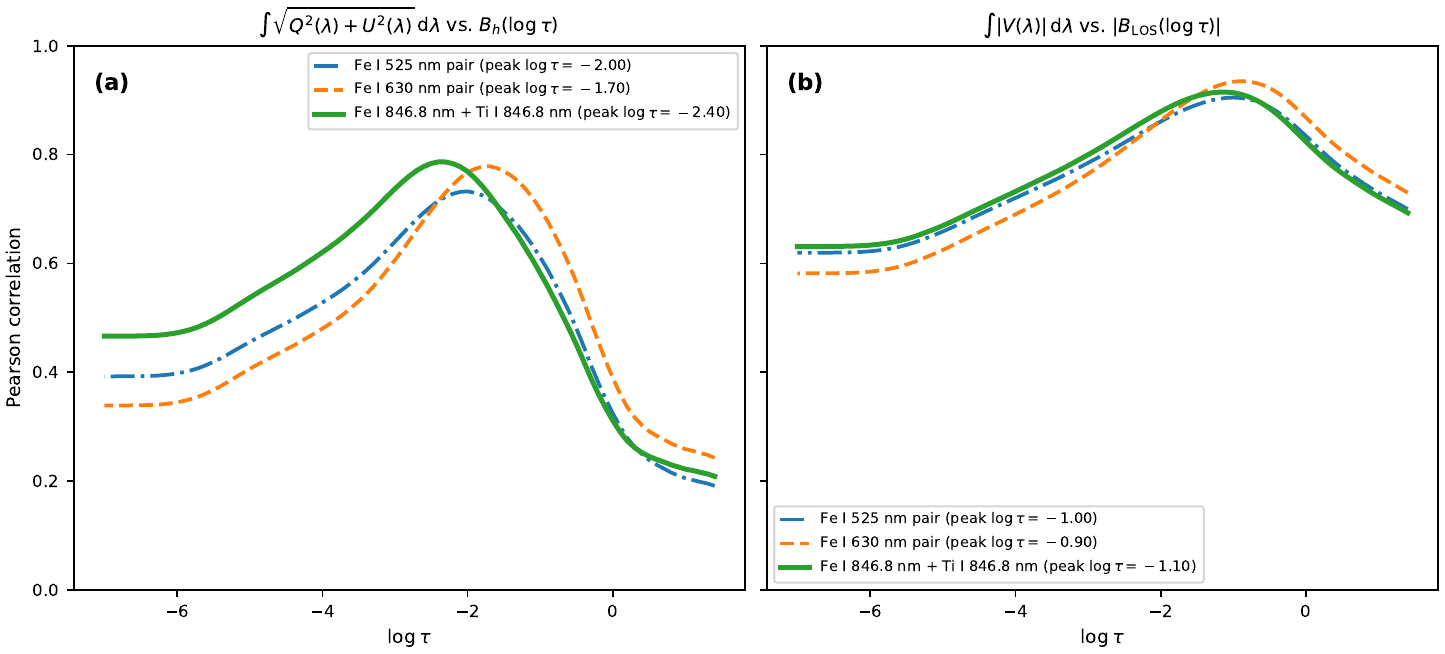}
    \caption{
Pearson correlation coefficients between synthesised polarisation observables and atmospheric magnetic-field components in the MURaM simulation as a function of optical depth. (a) Correlation between the wavelength-integrated linear polarisation,
$\int \sqrt{Q^2(\lambda)+U^2(\lambda)}\,{\rm d}\lambda$,
and the horizontal magnetic field component, $B_h$. (b) Correlation between the wavelength-integrated circular polarisation,
$\int |V(\lambda)|\,{\rm d}\lambda$,
and the absolute line-of-sight magnetic field component, $|B_{\rm LOS}|$.
Results are shown for the Fe\,{\sc i} 525\,nm pair, the Fe\,{\sc i} 630\,nm pair, and the blended Fe\,{\sc i} 846.8\,nm and Ti\,{\sc i} 846.8\,nm lines SCIP observed.
}\label{fig:appendix_correlation}
\end{figure*}

Figure~\ref{fig:appendix_correlation} shows that the wavelength-integrated linear polarisation reaches its maximum correlation with $B_h$ in the mid-to-upper photosphere for all three Fe\,{\sc i} spectral windows considered. The Fe\,{\sc i}\,846.8\,nm window exhibits the strongest correlation near $\log\tau=-2.4$, whereas the corresponding circular polarisation exhibits maximum correlation with $|B_{\rm LOS}|$ much lower in the atmosphere near $\log\tau=-1.1$. Similar behaviour is found for the Fe\,{\sc i}\,525\,nm and Fe\,{\sc i}\,630\,nm line pairs, although their linear-polarisation correlations peak at deeper layers ($\log\tau=-2.0$ and $-1.7$, respectively).

These empirical correlations demonstrate that, within the MURaM atmospheres employed here, the synthesised linear polarisation carries its strongest information on the horizontal magnetic field component in the mid-to-upper photosphere, whereas the circular polarisation is most strongly associated with the line-of-sight field around $\log\tau\approx-1$. Consequently, the differences between SINN and \textsc{DeSIRe} in the upper photosphere occur where the synthesised linear polarisation remains strongly correlated with $B_h$, while the correlation between circular polarisation and $|B_{\mathrm{LOS}}|$ is weaker than at deeper layers. This suggests that the two components may be constrained differently at these depths. Because the inferred median $|B_{\mathrm{LOS}}|$ is small in the upper photosphere, modest absolute differences in this denominator can produce substantial differences in $B_{\mathrm{ratio}}$, even when the inferred horizontal components remain similar.

\section{Inferred magnetic field component values}\label{sect:appendix_B}
Table~\ref{tab:magnetic_components} provides the median magnetic field strengths and components at $\log\tau=-1.0$ for the full field of view and for the strong- and weak-field populations shown in Fig.~\ref{fig:Bh_vs_Blos}. The close agreement between \textsc{DeSIRe} and SINN in the weak-field population is evident in both the absolute field components and $B_{\mathrm{ratio}}$.

\begin{table}
\centering
\caption{Median magnetic field properties inferred by \textsc{DeSIRe} and SINN at $\log\tau=-1$. Network and internetwork are defined from pixels satisfying both the LP and Stokes $V$ detection criteria at $4\sigma$. The upper quartile in Stokes $V$ amplitude defines network and the remaining pixels define internetwork. Full-FOV values impose no polarisation selection.}
\label{tab:magnetic_components}
\begin{tabular}{llrrrr}
\hline\hline
Region & Method & $B$ & $B_h$ & $|B_{\mathrm{LOS}}|$ & $B_{\mathrm{ratio}}$ \\
 & & [G] & [G] & [G] & \\
\hline
Full FOV & \textsc{DeSIRe} & 40 & 38 & 4.7 & 8.1 \\
 & SINN & 45 & 43 & 5.6 & 7.8 \\
\hline
Network & \textsc{DeSIRe} & 103 & 76 & 50 & 1.5 \\
 & SINN & 97 & 83 & 37 & 2.3 \\
\hline
Internetwork & \textsc{DeSIRe} & 64 & 63 & 7.0 & 9.0 \\
 & SINN & 61 & 60 & 6.5 & 9.2 \\
\hline
\end{tabular}
\end{table}

\section{Sensitivity to the detection threshold}\label{sect:appendix_C}

The analysis in Sect.~\ref{sect:ratio} uses a particular polarisation detection threshold to restrict the sample to pixels containing significant information on both magnetic field components. To test whether the reported component ratios depend on this particular choice, we repeated the complete analysis using thresholds of $4\sigma$, $5\sigma$, and $6\sigma$. At each threshold, the linear- and circular-polarisation masks were recalculated independently. The boundary between the strong- and weak-signal populations was then recomputed as the 75th percentile of the Stokes-\(V\) amplitude distribution within the jointly detected sample. The selections contain $30.2\%$, $19.8\%$, and $13.7\%$ of the field of view, respectively. Thus, the tested thresholds change the size of the selected population by more than a factor of two.

Table~\ref{tab:threshold} lists the inferred component ratios at \(\log\tau=-1.0\). For the weak-signal population, DeSIRe returns ratios of 8.97, 8.58, and 8.14 at \(4\sigma\), \(5\sigma\), and \(6\sigma\), respectively. The corresponding SINN ratios are 9.19, 9.49, and 9.64. The inferred horizontal dominance therefore remains stable despite the substantial change in the number of selected pixels.

The strong-population ratios decrease as the detection threshold is increased, from $1.52$ to $1.11$ for DeSIRe and from $2.27$ to $1.59$ for SINN. This behaviour is expected because the strong/weak boundary is recalculated at each threshold: progressively higher thresholds select populations with larger Stokes-$V$ amplitudes and hence stronger line-of-sight magnetic components. These tests demonstrate that the inferred horizontal dominance of the detected weak-field population is not an artefact of the particular threshold adopted. When the full field of view is included, the ratios predicted by both methods become slightly more horizontal.

\begin{table}
\centering
\caption{Sensitivity of the inferred $B_{\mathrm{ratio}}$ to the
polarisation-detection threshold at $\log\tau=-1.0$. The selected FOV
fraction refers to pixels satisfying both the linear- and
circular-polarisation criteria.}
\label{tab:threshold}
\begin{tabular}{ccrrrr}
\hline\hline
Thres. &
 FOV &
\multicolumn{2}{c}{Strong} &
\multicolumn{2}{c}{Weak} \\
\cline{3-4}\cline{5-6}
&
(\%) &
DeSIRe &
SINN &
DeSIRe &
SINN \\
\hline
No thres. & 100.0 & 2.28 & 3.11 & 10.81 & 9.68 \\
$4\sigma$ & 30.2 & 1.52 & 2.27 & 8.97 & 9.19 \\
$5\sigma$ & 19.8 & 1.33 & 1.97 & 8.58 & 9.49 \\
$6\sigma$ & 13.7 & 1.11 & 1.59 & 8.14 & 9.64 \\
\hline
\end{tabular}
\end{table}

\end{appendix}

\end{document}